\documentclass[aps,pra,superscriptaddress,amsmath,amssymb]{revtex4-1}

\usepackage{comment}
\usepackage{tikz}
\usetikzlibrary{trees}
\usetikzlibrary{decorations.pathmorphing}
\usetikzlibrary{decorations.markings}
\tikzset{
    photon/.style={decorate, decoration={snake,segment length=1.5mm}, draw=black},
    coulomb/.style={dotted},
    electron/.style={draw=black, postaction={decorate},
        decoration={markings,mark=at position .55 with {\arrow[draw=black]{>}}}}, 
    gluon/.style={decorate, draw=magenta,
        decoration={coil,amplitude=4pt, segment length=5pt}},
    boundelectron/.style={thick, double},
    transverse/.style={dashed}
}

\usepackage{graphicx}
\usepackage{dcolumn}
\usepackage{bm}
\usepackage{rotating}
\usepackage{dcolumn}

\newcolumntype{.}{D{.}{.}{8}}

\usepackage[utf8]{inputenc}
\usepackage{physics}
\usepackage{amsmath, amssymb}
\usepackage{tabularx,booktabs,array,dcolumn}
\usepackage{setspace}
\usepackage{vmargin}
\usepackage{xcolor}
\usepackage{graphicx,psfrag,subfigure}

\usepackage{float}
\usepackage[all]{xy}

\usepackage{bm}
\usepackage{mathtools}
\usepackage[english]{babel}
\usepackage{siunitx,letltxmacro}
\LetLtxMacro{\svqty}{\qty}
\usepackage{physics}
\LetLtxMacro{\qty}{\svqty}

\newcommand{\bos}[1]{\boldsymbol{#1}}
\newcommand{\mr}[1]{\mathrm{#1}}
\newcommand{\pd}[2]{\frac{\partial #1}{\partial #2}}

\newcommand{\bA}{\boldsymbol{A}}

\def\Eh{E_\mathrm{h}}

\def\iim{\mr{i}}

\def\pp{{++}}

\def\DC{\text{DC}}
\def\DCB{\text{DCB}}
\def\DCpB{{\text{DC}\langle \text{B}\rangle}}
\def\DCppB{{\text{DC}\mathcal{B}_2}}

\def\nnuc{N_\text{nuc}} 

\def\unittwo{1^{[2]}} 
\def\unitfour{1^{[4]}} 

\def\four{^{[4]}} 

\def\tT{\text{T}}

\def\bp{\bos{p}}
\def\br{\bos{r}}
\def\bs{\bos{s}}

\def\bsigma{\bos{\sigma}}

\def\mL{\Lambda}

\def\pp{{++}}

\def\som{Supplementary Material}

\def\texto{\text{o}}
\def\texte{\text{e}}

\usepackage[unicode]{hyperref}
\usepackage{soul}

\definecolor{ao}{rgb}{0.0, 0.5, 0.0}

\newcolumntype{d}[1]{D{.}{.}{#1}}

\usepackage[unicode]{hyperref}
\hypersetup{
   unicode=true,          
   plainpages=false,
   colorlinks=true,       
   linkcolor=blue,          
   citecolor=blue,        
}

\usepackage{url}
\begin{document}

\title{%
Relativistic two-electron atomic and molecular energies using \texorpdfstring{$LS$}{} coupling and double groups: role of the triplet contributions to singlet states 
}

\author{P\'eter Jeszenszki} 
\email{jeszenszki.peter@ttk.elte.hu}
\author{Edit M\'atyus} 
\email{edit.matyus@ttk.elte.hu}
\affiliation{ELTE, Eötvös Loránd University, Institute of Chemistry, 
Pázmány Péter sétány 1/A, Budapest, H-1117, Hungary}

\date{\today}

\begin{abstract}
\noindent %
The triplet contribution is computed to the 1 and 2\ $^1S^\texte_0$ states of the He atom, to the $1\ ^1S^\texte_0$ state of the Li$^+$ and Be${^{2+}}$ ions, and to the $X\ ^1\Sigma_\text{g}^+$ ground state of the H$_2$ molecule by extensive use of double-group symmetry (equivalent to $LS$ coupling for the atomic systems) during the course of the variational solution of the no-pair Dirac--Coulomb--Breit wave equation. 
The no-pair Dirac--Coulomb--Breit energies are converged within a sub-parts-per-billion relative precision 
using an explicitly correlated Gaussian basis optimized to the non-relativistic energies.
The $\alpha$ fine-structure constant dependence of the triplet sector contribution to the variational energy is $\alpha^4\Eh$ at leading order, in agreement with the formal perturbation theory result available from the literature.
\end{abstract}

\maketitle

\section{Introduction \label{sec:intro}}
\noindent%
Precision spectroscopy measurements and high-accuracy computations of atoms and molecules extend the boundaries of current scientific knowledge, which up to this point justify predictions of quantum electrodynamics (QED) \cite{salumbidesBoundsFifthForces2013,safronovaSearchNewPhysics2018,ahmadiCharacterization1S2S2018}. 
Regarding the theoretical predictions of atoms and molecules of light elements, the current state-of-the-art is based on computations within the non-relativistic quantum electrodynamics (nrQED) framework \cite{pachuckiNonrelativisticQEDApproach2004,pachuckiHigherorderEffectiveHamiltonian2005,patkosCompleteQuantumElectrodynamic2019,wehrliQEDEffectNuclear2021,earwoodRelativisticQEDCorrections2021}. In nrQED, well-converged non-relativistic states are considered as reference, and the relativistic and QED corrections are accounted for in a perturbative manner. The correction terms improve the non-relativistic energy according to an expansion for the $\alpha$ fine-structure constant and its nuclear-charge number multiple, $Z \alpha$, leading to an excellent agreement with the experimental results  \cite{puchalskiNonadiabaticQEDCorrection2019a,yerokhinAtomicStructureCalculations2021}. To maintain this high accuracy for increasing nuclear charge numbers, the perturbative treatment must be extended with (not-yet-available) `higher-order corrections' \cite{yerokhinTheoreticalEnergiesLowlying2010,yerokhinQEDCalculationsEnergy2022a}.

For high nuclear charges, the $1/Z$ expansion  \cite{shabaevTwotimeGreenFunction2002,mohrQuantumElectrodynamicsHigh1985,volotkaManyElectronQEDCorrections2014} can be successfully used by accounting for `all orders' of $Z \alpha$ in the reference state and treating the inter-particle correlations as perturbation. The medium-$Z$ region presents challenges for both the nrQED and the $1/Z$ approaches due to the competing importance of particle correlation and relativistic and QED contributions. A practical solution is offered by a combination of the two established methodologies (nrQED and $1/Z$ expansion) leading to a so-called `unified approach', in which a careful selection of $1/Z$ terms can complete the missing higher-order terms of the nrQED expansion \cite{drakeUnifiedRelativisticTheory1979,drakeTheoreticalEnergiesStates1988,yerokhinTheoreticalEnergiesLowlying2010,yerokhinElectroncorrelationEffectsFactor2017,yerokhinQEDCalculationsEnergy2022a}.

Inclusion of the relativistic `effects' in the zeroth-order state appears to be a promising alternative. 
For two electrons, the equal-time form \cite{sucherEnergyLevelsTwoElectron1958} of the Bethe--Salpeter equation \cite{betheQuantumMechanicsOne1977}  can be used as a starting point \cite{MaFeJeMa22}. 
This framework, in principle, allows us to develop perturbative corrections to the no-pair Dirac--Coulomb(--Breit) state incorporating both the particle correlation and relativistic effects in the zeroth-order description. The reference state
includes a partial resummation in $Z\alpha$, which is expected to accelerate the convergence of the perturbative expansion.

We aim to compute highly accurate energy levels by solving the no-pair Dirac--Coulomb(--Breit) equation by explicitly considering the inter-particle correlations. 
According to the literature, the wave function could be expanded in a determinant basis set by considering the explicit correlation in a perturbative framework \cite{liRelativisticExplicitCorrelation2012,liuPerspectivesRelativisticQuantum2012,liuRelativisticExplicitCorrelation2017}, or it can be  expanded directly in an explicitly correlated basis set \cite{bylickiRelativisticHylleraasConfigurationinteraction2008,jeszenszkiAllorderExplicitlyCorrelated2021,ferencVariationalVsPerturbative2022}. 
In an explicitly correlated, relativistic framework, 
a crucial step is the projection onto the positive-energy space. It is different from determinant-based approaches, in which the positive-energy space can be identified according to the single-particle energies. 
Liu and co-workers proposed a `dual-basis' approach to describe the external space spanned by the auxiliary explicitly correlated basis functions   \cite{liRelativisticExplicitCorrelation2012,liuPerspectivesRelativisticQuantum2012,liuRelativisticExplicitCorrelation2017}. 
At the same time, for an explicitly correlated (finite) basis set, the single-particle energies (and states) are unknown. Bylicki, Pestka, and Karwowski proposed to use a complex scaling (CS or complex coordinate rotation, CCR) approach to identify the positive-energy space in an explicitly correlated framework, without relying on the knowledge of single-particle energies in the finite basis set \cite{pestkaApplicationComplexcoordinateRotation2006,pestkaComplexCoordinateRotation2007,bylickiRelativisticHylleraasConfigurationinteraction2008,jeszenszkiVariationalDiracCoulombExplicitly2022}. 
According to our recent experience for atoms and molecules of light elements \cite{jeszenszkiAllorderExplicitlyCorrelated2021,jeszenszkiVariationalDiracCoulombExplicitly2022,ferencVariationalVsPerturbative2022,MaFeJeMa22}, the positive-energy space can be practically selected to a very good approximation by introducing an energy cutoff on the non-interacting two-electron energies (`cutting projector').
For the medium-$Z$ range, an optimal combination of the small-angle CCR and the cutting projector ideas might lead to a `punching projector' approach \cite{jeszenszkiVariationalDiracCoulombExplicitly2022}, but a robust, practical implementation and use of such a punching projector requires further work.

In this paper, our previous work on the no-pair Dirac--Coulomb(--Breit) Hamiltonian using explicitly correlated Gaussian (ECG) basis functions \cite{jeszenszkiAllorderExplicitlyCorrelated2021,jeszenszkiVariationalDiracCoulombExplicitly2022,ferencBreitInteractionExplicitly2022a,ferencVariationalVsPerturbative2022} is complemented by an efficient inclusion of the triplet space in the basis set expansion by considering double group projection.

In principle, the variational procedure would ensure that the correct spatial symmetries are numerically prepared even if the underlying basis functions do not have these symmetries \cite{SuUsVa98}. According to numerical experience for the non-relativistic energies \cite{muoloExplicitlyCorrelatedGaussian2018a}, use of an ECG basis that is not adapted to the spatial symmetries of the system would result in a slow convergence with respect to the basis size, and the best achievable convergence of the energy within double precision (limited by near-linear dependencies and computational time) is insufficient for spectroscopic purposes.

The symmetry adaptation using double groups, developed in the present work, is necessary for tight convergence of the Dirac--Coulomb(--Breit) energy for two-electron ions (atom) and the ground electronic state of the H$_2$ molecule including the complete basis space of the relativistic theory.

\section{Variational relativistic  explicitly-correlated methodology \label{sec:method}}
\noindent %
The no-pair Dirac--Coulomb (DC) Hamiltonian can be written in the following explicit form for two electrons 
\begin{align}
  &\mathcal{H}_\mathrm{DC}^{++[16]} = \nonumber \\
  &{ \footnotesize
  \mL_{++}
  \left(%
    \begin{array}{@{} c@{}c@{}c@{}c @{}}
       V\unitfour+U \unitfour & 
       c \bsigma\four_2 \cdot \bp_2 & 
       c \bsigma\four_1 \cdot \bp_1 & 
       0\four \\
       c\bsigma\four_2 \cdot \bp_2 & 
       V\unitfour+(U - 2m_2c^2)\unitfour & 
       0\four & 
       c \bsigma\four_1 \cdot \bp_1 \\
       c\bsigma\four_1 \cdot\bp_1 & 
       0\four &
       V\unitfour+(U-2m_1c^2)\unitfour & 
       c \bsigma\four_2 \cdot \bp_2 \\
       0\four & 
       c \bsigma\four_1 \cdot \bp_1 &
       c \bsigma\four_2 \cdot \bp_2 & 
       V\unitfour+(U-2m_{12}c^2)\unitfour \\
    \end{array}
  \right)
  \mL_{++}
  }
  \label{eq:fullHam}
\end{align}
with $m_{12}=m_1+m_2$, $\bp_i = -\iim(\pd{}{r_{ix}},\pd{}{r_{iy}},\pd{}{r_{iz}})$ ($i=1,2$),
$\bsigma\four_1=(\sigma_x\otimes\unittwo,\sigma_y\otimes\unittwo,\sigma_z\otimes\unittwo)$
and
$\bsigma\four_2=(\unittwo\otimes \sigma_x,\unittwo\otimes\sigma_y,\unittwo\otimes\sigma_z)$, where $\sigma_x,\sigma_y,$ and $\sigma_z$ are the $2\times 2$ Pauli matrices,  $U=\sum_{i=1}^2 \sum_{a=1}^{\nnuc}q_i Z_a/|\bos{r}_i-\bos{R}_a|$ is the external Coulomb interaction energy due to the nuclei, and $V=q_1 q_2/r_{12}$ is the Coulomb interaction between the particles. We note that both one-particle Hamiltonians are shifted by $-2m_ic^2$ $(i=1,2)$ for computational convenience and for a straightforward matching with the non-relativistic energy scale. 

Furthermore, we can include the (instantaneous) Breit interaction to arrive at the no-pair Dirac--Coulomb--Breit (DCB) operator, 
\begin{align}
  \mathcal{H}_\mathrm{DCB}^{++[16]} 
  &=  
  \mathcal{H}_\mathrm{DC}^{++[16]} 
  + 
  \Lambda_{++} B^{[16]} \Lambda_{++}
  \label{eq:DCB}
\end{align} 
with
\begin{align}
  B^{[16]} &= D^{[4]} \otimes B^{[4]} \ , \\
  D^{[4]} 
  &= 
  \left(%
  \begin{array}{cccc}
     0 & 0 & 0 & 1 \\
     0 & 0 & 1 & 0 \\
     0 & 1 & 0 & 0 \\
     1 & 0 & 0 & 0 
  \end{array} \right) \ ,
\end{align}
and
\begin{align}
  B^{\four}
  = 
  -\frac{q_1q_2}{2}
    \sum_{i=1}^3\sum_{j=1}^3 
      \left[%
      \frac{\delta_{ij}}{r_{12}}+\frac{1}{2} \left\lbrace \grad_{1_i}\grad_{2_j}r_{12} \right\rbrace 
      \right]
      \sigma^{[4]}_{1_i} \sigma^{[4]}_{2_j} \ .
\end{align}

The $\Lambda^{++}$ term is the positive-energy projector, which is commonly defined with respect to single-particle operators. Due to the explicitly correlated basis ({\it vide infra}) used in the present work, it is more convenient to define the projector using the positive-energy solution of the two-particle Hamiltonian in Eq.~(\ref{eq:fullHam}) without explicit particle-particle interaction ($V=0$ and  $B\four =0 \four$).

The relativistic energies and wave functions are obtained as the eigenvalues and eigenfunctions of the no-pair DC(B) Hamiltonian, 
\begin{align}
    \mathcal{H}_\mathrm{DC(B)}^{++[16]} \left| \Psi^{(16)} \right \rangle = E_\mathrm{DC(B)}^{++} \left| \Psi^{(16)} \right \rangle \ . \label{eigenvaleq} 
\end{align}
The sixteen-dimensional wave function can be conveniently written in a  block-wise spinor structure form,
\begin{align}
\left|  \Psi^{(16)} \right \rangle  = 
\begin{bmatrix}
\left| \psi^{\mathrm{l l}(4)} \right \rangle \\[0.2cm]
\left| \psi^{\mathrm{l s}(4)} \right \rangle \\[0.2cm]
\left| \psi^{\mathrm{s l}(4)} \right \rangle\\[0.2cm]
\left| \psi^{\mathrm{s s}(4)} \right \rangle 
\end{bmatrix} 
\hspace{1cm} \mbox{with} \hspace{1cm}
  \left| \psi^{\lambda_1 \lambda_2 (4)} \right \rangle  =
\begin{bmatrix}
\left| \psi^{\lambda_1 \lambda_2}_{\uparrow \uparrow} \right \rangle   \\[0.2cm]
\left| \psi^{\lambda_1 \lambda_2}_{\uparrow \downarrow} \right \rangle  \\[0.2cm]
\left| \psi^{\lambda_1 \lambda_2}_{\downarrow \uparrow} \right \rangle \\[0.2cm]
\left| \psi^{\lambda_1 \lambda_2}_{\downarrow \downarrow} \right \rangle   
\end{bmatrix} \hspace{0.5cm} (\lambda_1,\lambda_2=\text{l},\text{s}) \ , 
\label{spinor}
\end{align}
where `$\mathrm{l}$' and `$\mathrm{s}$' indicates the large and small components, $\uparrow$ and $\downarrow$ stand for the $+1/2$ and $-1/2$ spin projection of the electron. 

In order to solve Eq.~(\ref{eigenvaleq}), $\left|  \Psi^{(16)} \right \rangle$ is expanded in a basis of sixteen-component functions,
\begin{align}
   \left|  \Psi^{(16)} \right \rangle  
   &= 
   P_G^{\zeta {[16]}} 
   \left(%
     \sum_{i=1}^{N_\text{b}} \sum_{q=1}^{16} 
       c_{iq} 
       \mathcal{A}^{[16]} X^{[16]} | \Phi_{iq}^{(16)} \rangle 
   \right) \ , 
   \label{eq:ansatz}
\end{align}
where $c_{iq}$ is the linear expansion coefficient, $\mathcal{A}^{[16]}$ is the anti-symmetrization operator for the two electrons \cite{jeszenszkiVariationalDiracCoulombExplicitly2022}, and the $P_G^{\zeta [16]}$ is a projector corresponding to the $\zeta$ irreducible representation (irrep) of the $G$ point group (Sec.~\ref{sec:doublegroup}).  The $X^{[16]}$ matrix implements the restricted kinetic-balance condition \cite{kutzelniggBasisSetExpansion1984,sunComparisonRestrictedUnrestricted2011,jeszenszkiVariationalDiracCoulombExplicitly2022,ferencBreitInteractionExplicitly2022a},
 \begin{align}
      X=X^{[16]}  
      &=
      \text{diag}\left(%
      1\four, 
      \frac{\bsigma_2\four \bp_2 }{2m_2c},%
      \frac{\bsigma_1\four \bp_1 }{2m_1c},%
      \frac{\bsigma_1\four \bp_1 \bsigma_2\four \bp_2 }{4m_1m_2c^2}\right) \, .
      \label{eq:kinbal}
\end{align}
Henceforth, $X$ will be used instead of $X^{[16]}$ for brevity. 

The expression in the parenthesis on the right-hand side of Eq.~(\ref{eq:ansatz}) can already represent a correct wave function for relativistic computations, but the $P^{\zeta [16]}_G$ symmetry projector can accelerate the convergence with respect to the basis set size. 

A $| \Phi_{iq}^{(16)} \rangle$  sixteen-component basis function can be defined as 
\begin{align}
    | \Phi_{iq}^{(16)} \rangle  &= 1_q^{(16)}  \, | \Theta_i \rangle \  \label{eq:basis}
\end{align}
with $(1_q^{(16)})_p= \delta_{pq}$. 
$\Theta_i(\br)$ is a floating explicitly correlated Gaussian (fECG) function \cite{mitroyTheoryApplicationExplicitly2013,szalewiczExplicitlycorrelatedGaussianGeminals2010,suzukiStochasticVariationalApproach1998,matyusMolecularStructureCalculations2012a,matyusPreBornOppenheimerMolecular2019}, 
\begin{align}
   \Theta_i(\br) &= \exp \left[-\left( \br-\bs_i \right)^T \underline{\bos{A}}_i \left( \br - \bs_i \right)\right] \ ,
   \label{eq:ecg}
\end{align}
where the $\underline{\bos{A}}_i=\bA_i\otimes 1^{[3]}$  positive-definite exponent matrix with $\bos{A}_i\in\mathbb{R}^{2\times 2}$ and the $\bs_i\in\mathbb{R}^{3\times 2}$ `shift' vector are optimized variationally by minimization of the non-relativistic energy. 

In order to determine the linear combination coefficients in Eq.~(\ref{eq:ansatz}), we multiply Eq.~(\ref{eigenvaleq}) with $\langle \Phi^{(16)}_{jp}| X^\dagger \mathcal{A}^\dagger$  from the left
\begin{align}
  &
  \sum_{i=1}^{N_b} \sum_{q=1}^{16}  
  \left \langle %
    \Phi_{jp}^{(16)} \left| 
    X^\dagger \mathcal{A}^{[16]\dagger} %
    \mathcal{H}_{\mathrm{DC(B)}}^{++[16]} %
    \mathcal{P}_{G}^{\zeta [16]} \mathcal{A}^{[16]}  X 
    \right | \Phi_{iq}^{(16)} \right \rangle c_{iq}  
    = \nonumber \\ 
  & \hspace{5cm}  
  E_\mathrm{DC(B)}^{++}  
  \sum_{i=1}^{N_b} \sum_{q=1}^{16} 
  \left \langle %
    \Phi_{jp}^{(16)} \right. 
    \left| X^\dagger \mathcal{A}^{[16]\dagger} %
    P_{G}^{\zeta [16]} \mathcal{A}^{[16]} 
    X \right | \left. \Phi_{iq}^{(16)} %
  \right \rangle c_{iq}
  \ ,
  \label{eq:mxeigenalue0}
\end{align}
and then, rearrange this equation in order to arrive at matrix elements similar to our earlier work \cite{jeszenszkiAllorderExplicitlyCorrelated2021,jeszenszkiVariationalDiracCoulombExplicitly2022,ferencBreitInteractionExplicitly2022a,ferencVariationalVsPerturbative2022}, in which the $X$ kinetic balance was used in the sense of a metric. 
To arrive at this form, we use the commutation relations
\begin{align}
  \mathcal{A}^{[16]}\mathcal{H}_{\mathrm{DC(B)}}^{++[16]}
  =
  \mathcal{H}_{\mathrm{DC(B)}}^{++[16]}\mathcal{A}^{[16]} \ ,
  \hspace{2cm}
   \mathcal {A}^{[16]} X =  X \mathcal {A}^{[16]} \ , \\
    \mathcal{P}_{G}^{\zeta [16]}\mathcal{A}^{[16]}
  =
  \mathcal{A}^{[16]}\mathcal{P}_{G}^{\zeta [16]} \ ,
  \hspace{2cm}
  \mathcal{P}_{G}^{\zeta [16]} X
  =
  X \mathcal{P}_{G}^{\zeta [16]} \ ,
  \end{align}
and the idempotency of $\mathcal{A}^{[16]}=(1^{[16]}-P^{[16]}_{12})/2$ \cite{jeszenszkiVariationalDiracCoulombExplicitly2022}, and  write the operators of the left-hand side of Eq.~(\ref{eq:mxeigenalue0}) as
\begin{align}
  &X^\dagger \mathcal{A}^{[16]\dagger} %
  \mathcal{H}_{\mathrm{DC(B)}}^{++[16]} %
  \mathcal{P}_{G}^{\zeta [16]} \mathcal{A}^{[16]}  X 
  \nonumber \\
  =\ &
  X^\dagger %
  \mathcal{H}_{\mathrm{DC(B)}}^{++[16]} %
  \mathcal{A}^{[16]\dagger}
  \mathcal{P}_{G}^{\zeta [16]} \mathcal{A}^{[16]}  X
  \nonumber \\
  =\ &
  X^\dagger %
  \mathcal{H}_{\mathrm{DC(B)}}^{++[16]} %
  \mathcal{A}^{[16]\dagger}\mathcal{A}^{[16]} 
  \mathcal{P}_{G}^{\zeta [16]}  X     
  \nonumber \\
  =\ &
  X^\dagger %
  \mathcal{H}_{\mathrm{DC(B)}}^{++[16]} %
  \mathcal{A}^{[16]} 
  \mathcal{P}_{G}^{\zeta [16]}  X     
  \nonumber \\
  =\ &
  X^\dagger %
  \mathcal{H}_{\mathrm{DC(B)}}^{++[16]} %
  X X^{-1}
  \mathcal{A}^{[16]} 
  \mathcal{P}_{G}^{\zeta [16]}  X       
  \nonumber \\
  =\ &\ 
  X^\dagger %
  \mathcal{H}_{\mathrm{DC(B)}}^{++[16]} %
  X 
  \mathcal{A}^{[16]}  \mathcal{P}_{G}^{\zeta [16]}  X^{-1}X
  \nonumber \\
  =\ &
  \mathcal{H}_{\mathrm{DC(B)}X}^{++[16]} %
  \mathcal{A}^{[16]} 
  \mathcal{P}_{G}^{\zeta [16]} \\
  =\ &
  \mathcal{H}_{\mathrm{DC(B)}X}^{++[16]} %
  \mathcal{P}_{G}^{\zeta [16]}
    \mathcal{A}^{[16]}
\end{align}
where the transformed Hamiltonian appears,
\begin{align}
  \mathcal{H}_{\mathrm{DC(B)}X}^{++[16]}
  =
  X^\dagger \mathcal {H}_\mathrm{DC(B)}^{++[16]} X \ .
  \label{eq:mxeigenalueaux}    
\end{align}
For the operators on the right-hand side of Eq.~(\ref{eq:mxeigenalue0}), we have
\begin{align}
  &X^\dagger \mathcal{A}^{[16]\dagger} %
  \mathcal{P}_{G}^{\zeta [16]} \mathcal{A}^{[16]}  X 
  \nonumber \\
  =\ &
  X^\dagger 
  \mathcal{P}_{G}^{\zeta [16]} \mathcal{A}^{[16]}  X \\
  =\ &
  X^\dagger X 
  \mathcal{P}_{G}^{\zeta [16]} \mathcal{A}^{[16]} 
\end{align}
Hence, we can write the matrix eigenvalue equation, Eq.~(\ref{eq:mxeigenalue0}), in a compact form as
\begin{align}
  &\sum_{i=1}^{N_b} \sum_{q=1}^{16}  \left \langle \Phi_{jp}^{(16)} \left| \mathcal{H}_{\mathrm{DC(B)}X}^{++[16]} \mathcal{A}^{[16]} \mathcal{P}_{G}^{\zeta [16]} \right | \Phi_{iq}^{(16)} \right \rangle c_{iq}  = 
  E_\mathrm{DC(B)}^{++}  \sum_{i=1}^{N_b} \sum_{q=1}^{16} \left \langle \Phi_{jp}^{(16)} \right. \left|  X^\dagger X  \mathcal{A}^{[16]}P_{G}^{\zeta [16]} \right | \left. \Phi_{iq}^{(16)} \right \rangle c_{iq}
  \ .
  \label{eq:mxeigenalue}
\end{align}

Similarly to our earlier work \cite{jeszenszkiAllorderExplicitlyCorrelated2021,ferencBreitInteractionExplicitly2022a,ferencVariationalVsPerturbative2022}, the Breit interaction can be included not only variationally, and it was also considered within a perturbative framework in addition to the no-pair Dirac--Coulomb reference state. 
In what follows, $\DCpB$\ and  $\DCppB$\ label the sum of the no-pair Dirac--Coulomb energy with up to first- and second-order perturbative Breit corrections,
\begin{align}
  E_{\DCpB,k}^{++} &=   E_{\DC,k}^{++} \  + \ \left \langle \Psi_{\DC,k}^{(16)} \left| B^{[16]} \right| \Psi_{\DC,k}^{(16)} \right \rangle  
\end{align}
and
\begin{align}
  E_{\DCppB,k}^{++}
  =
  E_{\DCpB,k}^{++} \ + \ 
    \sum_{n \neq 0} 
    \frac{%
      \abs{%
        \langle%
          \Psi^{\pp}_{\text{DC},k} 
          | X^\dagger B^{[16]} X | 
          \Psi^{{\pp}}_{\text{DC},n}
        \rangle
      }^2
    }{%
      E^{\pp}_{\text{DC},k}-E^{\pp}_{\text{DC},n}
    } 
  \; ,
\end{align}
respectively, where $n$ goes through the positive-energy space excluding the reference state, $\left | \Psi_{\DC,k}^{++}\right \rangle $.

\section{Implementation of the double-group symmetry \label{sec:doublegroup}}
\noindent%
The DCB Hamiltonian commutes with the point-group symmetry operators, which act on the coordinate and spin space and are collected in the so-called double group. For two electrons, the double point group is isomorphic to the point group (commonly used in the non-relativistic theory).

The projector to the $\zeta$ irrep of a finite $G$ double point group \cite{dyallIntroductionRelativisticQuantum2007,bunkerMolecularSymmetrySpectroscopy1998} can be expressed as a linear combination of the symmetry operators, $\mathcal{O}^{[16]}$, with the characters, $\chi_{G\mathcal{O}}^\zeta$, 
\begin{align}
    P_{G}^{\zeta[16]} 
    = 
    \sum_{\mathcal{O} \in G} 
    \chi_{G\mathcal{O}}^\zeta 
    \mathcal{O}^{[16]} \ .  \label{eq:projsym} 
\end{align}
For continuous groups, in particular, the $O(3)$ and $D_{\infty \text{h}}$ relevant for the numerical applications presented in this work, the summation is `replaced' with integration for the exact projector. To avoid cumbersome numerical integrals (or very lengthy analytic expressions), we use instead of the continuous group a conveniently chosen finite subgroup (explained in the \som\ in detail), for which the projection amounts to performing a finite summation, as shown in Eq.~(\ref{eq:projsym}).

Since we construct a basis representation for the relativistic problem using basis sets optimized for non-relativistic states with some point-group symmetry and some spin quantum numbers, the $\mathcal{O}^{[16]}$ spin-spatial symmetry operator is conveniently written as a product of an $O$ spatial and an $\mathbb{O}^{[16]}$ spin operator,
\begin{align}
  \mathcal{O}^{[16]}
  =
  O \mathbb{O}^{[16]}
  = 
  \mathbb{O}^{[16]} O \ .
  \label{eq:lsoper}
\end{align}
The effect of the $O$ spatial symmetry operator on the spatial ECG function, Eq.~(\ref{eq:ecg}), can be accounted for by the transformation of the $\bos{s}_i$ shift vector (see for example, Refs.~\cite{matyusMolecularStructureCalculations2012a,matyusPreBornOppenheimerMolecular2019})
    \begin{align}
  O \Theta_i(\br)
  &=
  O
  \exp\left[%
    \left(\br- \bs_i\right)^\tT 
    \underline{\bA}_i
    \left(\br-\bs_i\right)
  \right] 
  \nonumber \\
  &=
  \exp\left[%
    \left(\br- \underbrace{O \bs_i}_{\bs_i^{O}} \right)^\tT 
    \underbrace{O^{-\tT} \underline{\bA}_i O^{-1}}_{\underline{\bA}_i} 
    \left(\br-\underbrace{O \bs_i}_{\bs_i^{O}} \right)
  \right]  
  \nonumber \\
  &=
  \exp\left[%
    \left(\br- \bs_i^{O} \right)^\tT
    \underline{\bA}_i
    \left(\br-\bs_i^{O} \right)
  \right] \ ,
  \label{eq:spatialECGtransform}
\end{align}
where we exploited the $\underline{\bA}_i=\bA_i\otimes \bos{1}_3$ structure.
The corresponding $\mathbb{O}$ operator in Eq.~(\ref{eq:lsoper}) transforms the spinor functions and its explicit form was discussed in Ref.~\cite{jeszenszkiVariationalDiracCoulombExplicitly2022}.

Furthermore, it is advantageous to make the spatial and the spin terms explicit also in the double-group projector,  Eq.~(\ref{eq:projsym}), with relevance for the present applications, 
\begin{align}
    P_{G}^{\zeta[16]} 
    = 
    \sum_{%
      \stackrel{\forall \zeta',\zeta'':}%
      {\zeta=\zeta' \otimes \zeta''}
    } \, \, \, \, 
    \sum_{%
      \stackrel{\forall \mathbb{O}^{[16]},O}%
      {\mathcal{O}^{[16]}=\mathbb{O}^{[16]}O}
    }
    \chi_{G\mathbb{O}}^{\zeta'} \mathbb{O}^{[16]} \, \chi_{GO}^{\zeta''} 
     O \ ,  \label{eq:projsymsep} 
\end{align}
where we can exploit the fact that the double group is isomorphic with the spin symmetry and the spatial symmetry groups for all applications presented in this paper (Secs.~\ref{sec:numres}), and for simplicity, we use the same $G$ and irrep labels in all three cases. Hence, the summation for $\zeta'$ and for $\zeta''$ (corresponding to the operations $\mathbb{O}^{[16]}$ and $O$) run through the irreps for which $\zeta' \otimes \zeta''=\zeta$ (corresponding to $\mathbb{O}^{[ 16 ]}O=\mathcal{O}^{[16]}$).  
In this case, the double-group character is obtained as the product of the characters of the spin and spatial representations, $\chi_{G\mathcal{O}}^{\zeta}=\chi_{G\mathbb{O}}^{\zeta'}\chi_{GO}^{\zeta''}$. 

We write the two-particle basis functions as a product of a (two-particle) spatial function and a (two-particle) spinor function. Instead of using elementary spin functions, we use $S^2$ and $S_z$ eigenfunctions. Furthermore, for atoms (homonuclear diatomics) the $L^2$ ($L_z$)  and parity eigenfunctions are conveniently used for the spatial part. This basis choice is known as $LS$ coupling. So, instead of using elementary spin functions to construct the 16-component Dirac spinor, it is better to use $S^2$ eigenfunctions, which are singlet and triplet eigenfunctions for two electrons.

For later convenience, we do not consider $S_z$ eigenfunctions (drop the $M_S$ quantum number) and instead use the $x$, $y$, and $z$ representation for the $\Sigma_{1,\alpha}$ triplet subspace, 
\begin{align}
  \left(%
  \begin{array}{c}
    \Sigma_{1,y}    \\[0.2cm]
    \Sigma_{1,z}   \\[0.2cm]
    \Sigma_{0,0}  \\[0.2cm]
    \Sigma_{1,x}    
  \end{array} \right) 
  &= 
  \underbrace{%
    \frac{1}{\sqrt{2}} 
    \left(%
    \begin{array}{cccc}
      \iim & \hspace{0.3cm}0 &  \hspace{0.3cm}0 & -\iim \\
      0 & \hspace{0.3cm}\iim &  \hspace{0.3cm}\iim & \hspace{0.3cm}0 \\
      0 & \hspace{0.3cm}1 & -1 & \hspace{0.3cm}0 \\
      1 & \hspace{0.3cm}0 &  \hspace{0.3cm}0 & \hspace{0.3cm}1
   \end{array}
   \right) }_{%
    \mathcal{U}^{[4]}} 
    \left(%
      \begin{array}{c}
        \Sigma_{\uparrow \uparrow}   \\[0.2cm]
        \Sigma_{\uparrow \downarrow}   \\[0.2cm]
        \Sigma_{\downarrow \uparrow} \\[0.2cm]
        \Sigma_{\downarrow \downarrow}    
     \end{array} 
    \right)  \label{eq:Txyzdef}
\end{align}
with
\begin{align}
  \Sigma_{\uparrow \uparrow} 
  = 
  \left( \begin{array}{c}
    1 \\ 0 \\ 0 \\ 0 \\
  \end{array}\right) \ , \hspace{1cm} 
  \Sigma_{\uparrow \downarrow} 
  = 
  \left( \begin{array}{c}
    0 \\ 1 \\ 0 \\ 0 \\
  \end{array}\right) \ , \hspace{1cm}
  \Sigma_{\downarrow \uparrow} 
  = 
  \left( \begin{array}{c}
    0 \\ 0 \\ 1 \\ 0 \\
  \end{array}\right) \ , \hspace{1cm}
  \Sigma_{\downarrow \downarrow} 
  = 
  \left( \begin{array}{c}
    0 \\ 0 \\ 0 \\ 1 \\
  \end{array}\right) \ ,
\end{align}
corresponding to all $(\lambda_1,\lambda_2)$ blocks of Eqs.~(\ref{spinor}).

This representation, in comparison with an elementary spin representation, is advantageous, since the $\Sigma_{0,0}$ singlet function transforms according to the totally symmetric irrep, while the
$\Sigma_{1,x}, \Sigma_{1,y},$ and $\Sigma_{1,z}$
triplet functions transform according to the $R_x$, $R_y$, and $R_z$ rotation operators about the $x$, $y$, and $z$ axes 
of the symmetry group, $G$ \cite{dyallIntroductionRelativisticQuantum2007}. 

During the course of the numerical computations presented in this work, we always used the $D_{2\text{h}}$ double-point group (a subgroup of both the $O$(3) and $D_{\infty \text{h}}$ groups relevant for the atomic and homonuclear diatomic applications). In this double-point group, the $\Sigma_{0,0}$, $\Sigma_{1,x}$, $\Sigma_{1,y}$, and $\Sigma_{1,z}$  spin functions transform according to the $A_{\text{g}}$, $B_{3\text{g}}$, $B_{2\text{g}}$, and $B_{1\text{g}}$ irreps (Table S2), respectively.

The matrix representation of the Hamiltonians, Eqs.~(\ref{eq:fullHam})--(\ref{eq:DCB}), corresponding to an elementary spinor basis, can be also transformed to our singlet-triplet spinor basis using the $\mathcal{U}^{[4]}$ matrix in Eq.~(\ref{eq:Txyzdef}), 
\begin{align}
    \mathcal{H'}_{\lambda_1 \lambda_2}^{[4]}
    &= 
    \mathcal{U}^{[4]} \mathcal{H}_{\lambda_1 \lambda_2}^{ [4]} \mathcal{U}^{\dagger [4]} 
    \nonumber \\
    &=
    \left(%
    \begin{array}{cccc}
         H'_{\lambda_1 \lambda_2,T_y,T_y} & H'_{\lambda_1 \lambda_2,T_y,T_z}   & H'_{\lambda_1 \lambda_2,T_y,S}   & H'_{\lambda_1 \lambda_2,T_y,T_x}  \\
         H'_{\lambda_1 \lambda_2,T_z,T_y}  & H'_{\lambda_1 \lambda_2,T_z,T_z}   & H'_{\lambda_1 \lambda_2,T_z,S}   & H'_{\lambda_1 \lambda_2,T_z,T_x}  \\
         H'_{\lambda_1 \lambda_2,S,T_y}  & H'_{\lambda_1 \lambda_2,S,T_z}   & H'_{\lambda_1 \lambda_2,S,S}   & H'_{\lambda_1 \lambda_2,S,T_x}  \\
         H'_{\lambda_1 \lambda_2,T_x,T_x}  & H'_{\lambda_1 \lambda_2,T_x,T_z}   & H'_{\lambda_1 \lambda_2,T_x,S}   & H'_{\lambda_1 \lambda_2,T_x,T_x}  
    \end{array} \right)  \; , 
    \nonumber \\
    &\quad\quad\quad \lambda_1,\lambda_2=\text{l},\text{s} \; .
    \label{transfromedH}
\end{align}

We note that every transformation is carried out block-wise ($4\times 4$) for every $(\lambda_1,\lambda_2)$ block (large-large, large-small, small-large, and small-small). Using the spin-spatial factorized form of the projector, Eq.~(\ref{eq:projsymsep}),  the spin `part' can be evaluated `trivially', since in the $D_{\text{2h}}$ group, the effect of the $\chi_{g\mathbb{O}}^{\zeta'} \mathbb{O}^{[16]}$ operation is identity on the $\Sigma_{0,0}$, $\Sigma_{1,x}$, $\Sigma_{1,y}$, as well as $\Sigma_{1,z}$ functions in all $\zeta$ irreps. 
Hence, the effect of the projector, Eqs.~(\ref{eq:projsym}) and (\ref{eq:lsoper}), is evaluated as 
\begin{align}
    P_G^{\zeta [16]}  
    \left| \Sigma^{(16)}_{\bar{\zeta} q} \right \rangle \left | \Theta_i \right \rangle 
    &=    
    \underbrace{\sum_{\zeta'}\sum_{\zeta''}}_{\zeta=\zeta' \otimes \zeta''} \, \, \, \,
    \underbrace{\sum_{\mathbb{O}^{[16]}}\sum_{O}}_{G\ni\mathcal{O}^{[16]}=\mathbb{O}^{[16]}O}
    \underbrace{ \chi_{G\mathbb{O}}^{\zeta'} \mathbb{O}^{[16]}  \left| \Sigma^{(16)}_{\bar{\zeta} q} \right \rangle}_{\delta_{\zeta' \bar{\zeta}} \left| \Sigma^{(16)}_{\bar{\zeta} q} \right \rangle} \, \chi_{GO}^{\zeta''} 
    \ O \ \left | \Theta_i \right \rangle \\
%
    &=\left| \Sigma^{(16)}_{\bar{\zeta} q} \right \rangle   
     \sum_{\stackrel{O}{G \ni\mathcal{O}^{[16]}=\mathbb{O}^{[16]}O}} \chi_{GO}^{\bar{\zeta}''} \ O \ \left | \Theta_i \right \rangle 
     \quad \text{with} \quad 
     \zeta=\bar{\zeta} \otimes \bar{\zeta}'' \; .
\end{align}

Thus, in short, 
since we used the $\Sigma_{0,0}$ and $\Sigma_{1,\alpha}\ (\alpha=x,y,z)$ spin representation, 
we had to perform only the (spatial) projection for the fECG functions (presented in detail in the \som).

\section{Numerical results \label{sec:numres}}
\noindent%
We focus on (compounds of) light elements, for which a variationally optimized non-relativistic basis set parameterization can provide highly accurate relativistic energies \cite{jeszenszkiAllorderExplicitlyCorrelated2021,jeszenszkiVariationalDiracCoulombExplicitly2022,ferencBreitInteractionExplicitly2022a,ferencVariationalVsPerturbative2022}. 
So, the computational strategy in this work was 
(1) variational optimization of the non-relativistic energy for both the singlet and the triplet states, and 
(2) coupling these basis states in Dirac relativistic computations. 
The algorithm is implemented in the QUANTEN (QUANTum mechanical computations for Electrons and atomic Nuclei) progam package, an in-house developed molecular physics platform  \cite{matyusPreBornOppenheimerMolecular2019,ferencNonadiabaticMassCorrection2019,ferencComputationRovibronicResonances2019,ferencNonadiabaticRelativisticLeadingOrder2020,MaCa21,IrJeMaMaRoPo21,jeszenszkiInclusionCuspEffects2021,jeszenszkiAllorderExplicitlyCorrelated2021,jeszenszkiVariationalDiracCoulombExplicitly2022,ferencBreitInteractionExplicitly2022a,ferencVariationalVsPerturbative2022,MaFe22nad,FeMa22bethe}.

For the systems studied in this work \emph{(vide infra)} and a (sub) parts-per-billion (ppb) relative precision in the energy, in most of the cases,  double precision arithmetic provided sufficient numerical stability, with a few exceptions noted in the text (for which higher, 10-byte reals or 16-byte reals (quadruple) precision arithmetic was used). 

The numerical results reported in this work were obtained with the `cutting' projector \cite{jeszenszkiVariationalDiracCoulombExplicitly2022}. Up to medium basis set sizes, some of the results were checked also with the CCR projector \cite{jeszenszkiVariationalDiracCoulombExplicitly2022}. For the largest basis sets, the CCR computation became numerically unstable (and require further, extensive computations with increased precision arithmetic).

Throughout this work hartree atomic units are used, and the reported energies correspond to the speed of light with $c=\alpha^{-1}a_0\Eh/\hbar$ and $\alpha^{-1}=137.$035 999 084 \cite{Codata2018Recommended}.

\subsection{Computations for helium-like systems \label{sec:heliumlike}}
 In this subsection, we first \emph{test} the methodology for the lowest-energy,
 2~$^3P^\text{o}_{0,1,2}$ manifold (corresponding to the $^{2S+1}L^p_J$ notation) of the helium atom (Sec.~\ref{sec:triplet}). 
 Then in Sec.~\ref{sec:singletatom}, we pinpoint the relativistic energy of the 1~$^1S_0$ ground-state electronic state of the He atom, the Li$^+$ and Be$^{2+}$ ions including the complete relativistic space.
 
 The relativistic eigenfunctions of these systems transform according to the $O$(3) double group, and the exact quantum numbers, $J$, $M_J$, and $p$, correspond to the squared total angular momentum $\bos{J}^2$, its projection, $J_z$, and space inversion, respectively.
 According to the methodological details (Secs.~\ref{sec:method} and \ref{sec:doublegroup}), we 
 have considered the construction of the relativistic basis set using the $LS$ coupling scheme, in which 
 we can exploit accurate, explicitly correlated representations of the non-relativistic states with $L$ (orbital) and $S$ (spin) quantum numbers. 
 Relativistic basis states in $LS$ coupling can be written according to the angular momentum coupling rules \cite{condonTheoryAtomicSpectra1991}, used with the KB metric (Sec.~\ref{sec:method}) for every $\lambda_1,\lambda_2$ block as
 \begin{align}
     \Phi_{J,M_J,\lambda_1,\lambda_2}^{(p)} 
     = 
     \sum_{M_L=-L}^L \sum_{M_S=-S}^S 
     \left \langle L,M_L,S,M_S|J,M_J \right \rangle 
     \varphi_{L,M_L}^{(p)} \Sigma_{S,M_S} \ ,
     \label{CGcoeff}
 \end{align}
where the non-vanishing $\left \langle L,M_L,S,M_S|J,M_J \right \rangle$ Clebsch--Gordan coefficients, satisfying the $|L-S|\leq J \leq |L+S|$ triangular inequality, indicate the (non-relativistic) spatial and spin functions that are necessary to have a complete relativistic basis. During the course of the numerical computations, we used the $D_{\text{2h}}$ group, a subgroup representation of $O$(3) (\emph{cf.} Table~S1 of the \som for the corresponding irreps).

\subsubsection{The lowest-energy \texorpdfstring{$^3P^\texto_{J}$}{} manifold of the helium atom \label{sec:triplet}}
\begin{figure}
    \centering
    \includegraphics[scale=0.7]{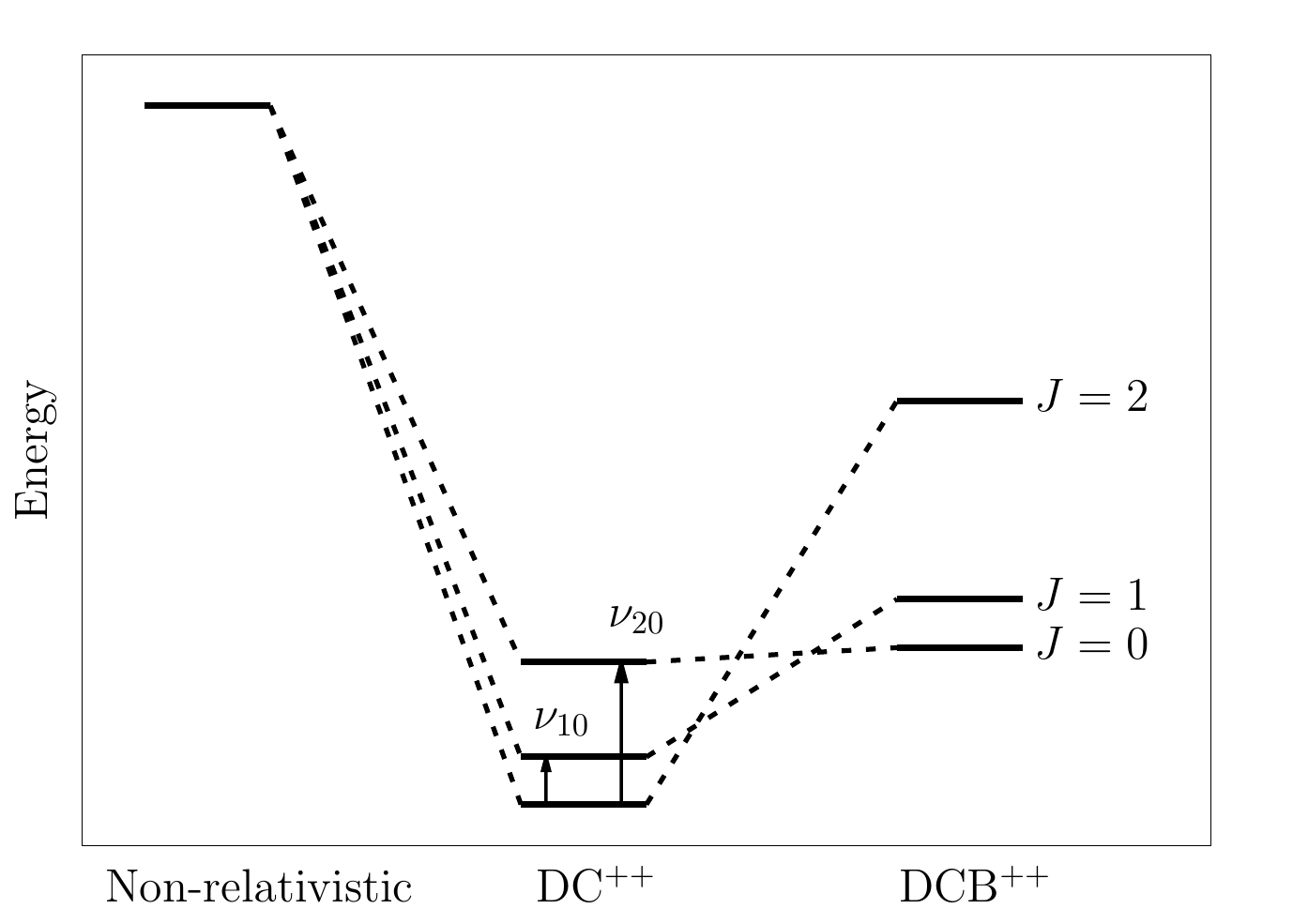}
    \caption{%
      Schematic diagram of the lowest-energy $^3P^\texto_J$ manifold of the helium atom.
    }
    \label{fig:3PHe}
\end{figure}
First, as a test of the implementation, we considered the fine-structure splitting (and the centroid energy) corresponding to the $^3P^\texto_J\ (J=0,1,2)$ manifold of the helium atom (`o' labels odd parity). 
This manifold consists of nine states, which are degenerate in the non-relativistic limit. 
In the DC(B) model, we obtain three levels, $^3P_{0}^\text{o}$, $^3P_{1}^\texto$, and $^3P_{2}^\texto$, with different energies for the different $J$ quantum numbers (Fig.~\ref{fig:3PHe}).

\vspace{0.5cm}
\paragraph{Symmetry considerations and technical details}
According to the $|L-S|\leq J \leq |L+S|$ triangular inequality (corresponding to Eq.~(\ref{CGcoeff})), 
the complete relativistic basis space, corresponding to the $^3P_J^\texto\ (J=0,1,2)$ manifold, can be constructed by the following types of functions (see also Eqs.~(\ref{eq:ansatz})--(\ref{eq:ecg}) and Eq.~(\ref{eq:Txyzdef})):
$(J=0,p=-1)$: $^3P^\texto$;
$(J=1,p=-1)$: $^1P^\texto$, $^3P^\texto$, $^3D^\texto$;
$(J=2,p=-1)$: $^3P^\texto$, $^1D^\texto$, $^3D^\texto$, $^3F^\texto$;
and we note that $^3S^\texto$ does not exist for two-electron systems \cite{UsSu00}.

Since, we consider the present fine-structure splitting computation as a first test of the implementation, we rely on physical-chemical intuition and perturbation theory knowledge \cite{drakeProgressHeliumFinestructure2002}, which tell us that the energy is dominated by contributions from $^3P^\texto$-type functions, \emph{i.e.,} the $\alpha^2\Eh$ part of the fine structure can be described using only the $^3P^\text{o}$-type functions. For $M_{J}=0$, these functions read as
  \begin{align}
    ^3P^\texto_0 : \, \, \, %
    \Phi_{0,0}^{(-1)}   
    &= 
    \frac{1}{\sqrt{6}} \left[ \sqrt{2}  \varphi_{1,-1}^{(-1)} \Sigma_{1,1} + \sqrt{2} \varphi_{1,1}^{(-1)} \Sigma_{1,-1} -  \varphi_{1,0}^{(-1)} \Sigma_{1,0} \right] \ , \\
    ^3P^\texto_1 : \, \, \,  %
    \Phi_{1,0}^{(-1)} 
    &=
    \frac{1}{\sqrt{2}} \left[ \varphi_{1,-1}^{(-1)} \Sigma_{1,1}  - \varphi_{1,1}^{(-1)} \Sigma_{1,-1}  \right] \ , \\
    ^3P^\texto_2 : \, \, \, %
    \Phi_{2,0}^{(-1)} 
    &= 
    \frac{1}{\sqrt{6}} \left[   \varphi_{1,1}^{(-1)} \Sigma_{1,1}+  \varphi_{1,-1}^{(-1)} \Sigma_{1,-1}+ \sqrt{2}\ \varphi_{1,0}^{(-1)} \Sigma_{1,0} \right]  \; .
  \end{align}

  According to the arguments made below Eq.~(\ref{eq:Txyzdef}), it is more convenient to replace the $L_z$ eigenfunctions with Cartesians ($x,y,z$) functions, 
  \begin{align}
    \varphi_{1,1}^{(-1)} &= \frac{1}{\sqrt{2}} \left[  \varphi_{1,x}^{(-1)}  + \iim \,  \varphi_{1,y}^{(-1)} \right] \ , \\
    \varphi_{1,-1}^{(-1)} &= \frac{1}{\sqrt{2}} \left[ \varphi_{1,x}^{(-1)}  - \iim \, \varphi_{1,y}^{(-1)}  \right] \ , \\ 
    \varphi_{1,0}^{(-1)}  &=  \varphi_{1,z}^{(-1)}  \ , 
  \end{align}
  which leads to
  \begin{align}
    \Phi_{0,0}^{(-1)} 
    &= 
    \frac{1}{\sqrt{3}} %
    \left[ 
      \varphi_{1x}^{(-1)} \Sigma_{1,x}  + \varphi_{1,y}^{(-1)} \Sigma_{1,y}   - \varphi_{1,z}^{(-1)} \Sigma_{1,z} 
    \right] \ , \label{phi00-1} \\
    \Phi_{1,0}^{(-1)}   
    &= 
    \frac{1}{\sqrt{2}} %
    \left[ \varphi_{1,x}^{(-1)}  \Sigma_{1,y}  - \varphi_{1,y}^{(-1)}  \Sigma_{1,x}  \right] \ , \label{phi10-1} \\
    \Phi_{2,0}^{(-1)} 
    &= 
    \frac{1}{\sqrt{6}} %
    \left[ \varphi_{1,x}^{(-1)} \Sigma_{1,x}  + \varphi_{1,y}^{(-1)} \Sigma_{1,y} + 2 \varphi_{1,z}^{(-1)}  \Sigma_{1,z}  \right]  \; .
    \label{phi20-1}
  \end{align}

The $\varphi_{1,\alpha}^{(-1)}\ (\alpha=x,y,z)$ functions were constructed in the $D_{2\text{h}}$ subgroup of $O$(3) with a special parametrization of the fECG functions, Eq.~(\ref{eq:ecg}).
First, the $\varphi_{1,z}^{(-1)}$ ($P_z^\texto$-type) spatial functions were represented with fECGs in which the $\bos{s}_i\in \mathbb{R}^{3\times2}$ shift vectors were restricted to the $z$ axis and were adapted to the $B_{1\text{u}}$ irrep of 
$D_{2\text{h}}$ (which corresponds to the $z$ linear operator in $D_{2\text{h}}$, see also the \som). The corresponding  $\varphi_{1,x}^{(-1)}$ and  $\varphi_{1,y}^{(-1)}$ ($P_x$- and $P_y$-type) functions were generated from the $\varphi_{1,z}^{(-1)}$ non-relativistically optimized functions by `moving' the $\bos{s}_i$ shift vectors from the $z$ to the $x$ and $y$ axes, respectively. These $\varphi_{1,x}^{(-1)}$ and  $\varphi_{1,y}^{(-1)}$ functions transform according to the $B_{3\text{u}}$ and $B_{2\text{u}}$ irreps (corresponding to the $x$ and $y$ linear operators) in $D_{2\text{h}}$.

By inspection of Eqs.~(\ref{phi00-1})--(\ref{phi20-1}), one can note that the $\Phi_{0,0}^{(-1)}$ and $\Phi_{2,0}^{(-1)}$ functions are obtained as a linear combination of $\varphi_{1x}^{(-1)} \Sigma_{1,x}$, $\varphi_{1,y}^{(-1)} \Sigma_{1,y}$, and $\varphi_{1,z}^{(-1)} \Sigma_{1,z}$, while the $\Phi_{1,0}^{(-1)}$ type of states can be obtained from different types of functions, $\varphi_{1,x}^{(-1)} \Sigma_{1,y}$ and $\varphi_{1,y}^{(-1)} \Sigma_{1,x}$. 

In practice, we performed two DC(B) computations. First, we included $\varphi_{1,x}^{(-1)}$, $\varphi_{1,y}^{(-1)}$, and $\varphi_{1,z}^{(-1)}$ type spatial functions combined with the $\Sigma_{1,x}$, $\Sigma_{1,y}$, and $\Sigma_{1,z}$ spinors, respectively, and obtained the $2\ ^3P^\texto_0$ and $2\ ^3P^\texto_2$ states as ground and excited states, respectively. Second, we used $\varphi_{1,x}^{(-1)}$, and $\varphi_{1,y}^{(-1)}$ type spatial functions combined with the $\Sigma_{1,y}$,  and $\Sigma_{1,x}$ spinors to obtain the $2\ ^3P^\texto_1$ state (as the ground state).

\vspace{0.5cm}
\paragraph{Discussion of the numerical results}
Table~\ref{tab:HetripletEdiff} collects numerical results for the fine-structure splittings, $\nu_{10}$ and $\nu_{20}$ defined as $\nu_{J0}=E_J-E_{0}\ (J=1,2)$, 
while Table~\ref{tab:Hetripletcentroid} shows the centroid energy, $E_\text{c}=(E_0+3E_1+5E_2)/9$, for $E_J$ of $2\ ^3P^\texto_J$ with $J=0,1,2$ total angular momentum quantum numbers. 
Regarding the convergence with respect to the number of fECG functions ($N$ in the tables),
the $E_\text{c}$ centroid energy typically converges slower than the $\nu_{10}$ and $\nu_{20}$ fine-structure splittings, but we think that all reported energies are converged within 0.1~n$\Eh$. (Further convergence tables are collected in the \som.)

In the tables, the no-pair DC(B) energies computed in this work are compared with the highly-accurate $\alpha^2\Eh$- and $\alpha^4\Eh$-order perturbative results compiled from the literature \cite{schwartzFineStructureHelium1964a,drakeProgressHeliumFinestructure2002}. 
The multi-configuration Dirac--Fock (MCDF) fine-structure splittings~\cite{boTheoreticalStudyInteresting2008} computed with the GRASP program~\cite{jonssonNewVersionGrasp2K2013} are also shown in Table~\ref{tab:HetripletEdiff}.
The MCDF fine-structure splitting is three orders of magnitude less converged, than ours, while the centroid energy (MCDF result not shown in Table~\ref{tab:Hetripletcentroid}) is estimated to be  four orders of magnitude less precise than ours.
We also note that the MCDF energies correspond to a (fundamentally) different projector from the $\Lambda_{++}$ used in this work (Sec.~\ref{sec:method}), but this difference could be seen in the numerical results only if the MCDF energies were better converged.

In what follows, we compare our energies with highly-accurate perturbative results up to $\alpha^4\Eh$ order compiled from the literature.
The $\nu_{10}$ and $\nu_{20}$ fine-structure splittings, are in an excellent agreement with the $\alpha^2\Eh$-order perturbative values (obtained from a precise expectation value of the `spin-dependent' Breit--Pauli Hamiltonian with well-converged non-relativistic wave functions), the deviation is 1.1--1.3~n$\Eh$ (Table~\ref{tab:HetripletEdiff}). 

The $\alpha^3\Eh$-order non-radiative (relativistic) corrections for the DC and DCB models is zero for triplet states due to the vanishing $\langle \delta(\bos{r}_{12})\rangle_\text{nr}$ expectation value.\cite{sucherEnergyLevelsTwoElectron1958,douglasQuantumElectrodynamicalCorrections1974}

Ref.~\cite{drakeProgressHeliumFinestructure2002} (following Ref.~\cite{douglasQuantumElectrodynamicalCorrections1974}) reports $\alpha^4\Eh$-order corrections. Inclusion of the $\alpha^4\Eh$-order non-radiative corrections in the perturbative result reduces the deviation of the no-pair variational and perturbation theory results (corresponding to the $^3P$ sector) to 0.1 and $-0.2$~n$\Eh$ for $\nu_{10}$ and $\nu_{20}$, respectively.
It is necessary to note that this $\alpha^4\Eh$-order non-radiative correction~\cite{drakeProgressHeliumFinestructure2002} includes also retardation effects and pair corrections, since the $\alpha^4\Eh$-order DC(B) relativistic contributions were not reported separately in Ref.~\cite{douglasQuantumElectrodynamicalCorrections1974} or \cite{drakeProgressHeliumFinestructure2002}.

Regarding the $E_\text{c}$ centroid energy (Table~\ref{tab:Hetripletcentroid}), the no-pair variational and $\alpha^2\Eh$-order perturbative energy differ by $-$10.0~n$E_\mathrm{h}$.  
This $-$10.0~n$\Eh$ deviation is in a close agreement with the full $\alpha^4\Eh$-order non-radiative correction (after cancelling the divergences with the some radiative terms) for the $^3P$-type basis states,
$E_Q+E_A'=-11.017$~n$\Eh$ \cite{pachuckiHeliumEnergyLevels2006}.

These results (fine-structure splittings and centroid energy) confirm, in addition to Ref.~\cite{ferencVariationalVsPerturbative2022} (achieved for singlet states), that our no-pair variational approach \cite{jeszenszkiAllorderExplicitlyCorrelated2021,jeszenszkiVariationalDiracCoulombExplicitly2022,ferencBreitInteractionExplicitly2022a,ferencVariationalVsPerturbative2022} is consistent with the established perturbative nrQED procedure, and the numerical results highlight that the pair corrections and retardation corrections are of the order of 1~n$\Eh$ for the studied ($Z=2$) system. 

The present \emph{test} computations with the no-pair variational methodology were carried out using
(the dominant) $^3P$-type functions to describe the energy levels of the $2\ ^3P_J$ manifold of the helium atom, and provide an additional test of the methodology and implementation to the singlet-state results \cite{jeszenszkiAllorderExplicitlyCorrelated2021,jeszenszkiVariationalDiracCoulombExplicitly2022,ferencBreitInteractionExplicitly2022a,ferencVariationalVsPerturbative2022}. 
As it will be demonstrated in the next section, we can include other `sectors' of the relativistic basis space in the variational procedure. 

For a $\ ^3P_J$ manifold, additional relevant functions are of type
$^1P^\texto$, $^1D^\texto$, $^3D^\texto$, and $^3F^\texto$. Variational inclusion of all these types of basis functions  would result in a very large basis space (and very large matrices) for the no-pair computation. 
For the particular case of the $2\ ^3P_J$ manifold, it would probably be more economical to account for the effect of these basis sectors as perturbative corrections to a no-pair DCB computation carried out in the $^3 P_J$ basis, which is left for future work.
We only note at this point that it is known from formal perturbation theory (nrQED) that these other symmetry sectors give contribution starting at the $\alpha^4\Eh$ order.
Ref.~\cite{drakeProgressHeliumFinestructure2002} (Table~6) reports the energy increments for every basis sectors to the fine-structure splitting, which sum to $\nu^{[4]}_{10,\text{DCB}}(^1P^\texto,^3D^\texto)=-1.006$~n$\Eh$ and 
$\nu^{[4]}_{20,\text{DCB}}(^1D^\texto,^3D^\texto,^3F^\texto)=-0.023$~n$\Eh$ with, the largest contribution, by far, from $^1P$. 
Regarding the centroid energy, according to Table~II of Ref.~\cite{pachuckiHeliumEnergyLevels2006}, the cumulative contribution of basis states other than $^3P$ amounts to $-0.478$~n$\Eh$.

\begin{table}[ht]
  \caption{%
  Fine-structure splitting, $\nu_{J0}=E_J-E_{0}\ (J=1,2)$ in n$\Eh$, for the 2~$^3P^\texto_J\ (J=0,1,2)$ manifold of the helium atom (see also Fig.~\ref{fig:3PHe}) obtained from no-pair Dirac--Coulomb(--Breit) computations and comparison with perturbative and multi-configuration Dirac--Fock (MCDF) results.
  (The centroid energy is reported in Table~\ref{tab:Hetripletcentroid}).
  \label{tab:HetripletEdiff}
  }
  \begin{tabular}{@{} l d{-1}d{-1}d{-1} d{-1}d{-1}d{-1} d{-1}d{-1} @{}}
    \hline\hline\\[-0.35cm]
    \multicolumn{1}{l}{$N$} &
    \multicolumn{1}{c}{$\nu_{10,\mathrm{DC}}^{++}$}  &
    \multicolumn{1}{c}{$\nu_{20,\mathrm{DC}}^{++}$}  &
    \multicolumn{1}{c}{$\nu_{10,\DCpB}^{++}$}  &
    \multicolumn{1}{c}{$\nu_{20,\DCpB}^{++}$} &
    \multicolumn{1}{c}{$\nu_{10,\DCppB}^{++}$}  &
    \multicolumn{1}{c}{$\nu_{20,\DCppB}^{++}$} &
    \multicolumn{1}{c}{$\nu_{10,\mathrm{DCB}}^{++}$}  &
    \multicolumn{1}{c}{$\nu_{20,\mathrm{DCB}}^{++}$} 
    \\
    \hline\\[-0.35cm]
    300	& 932.1 & 2795.5 & \hspace{-0.3cm}-4493.3 & \hspace{-0.3cm}-4846.0 & \hspace{-0.3cm}-4492.0 & \hspace{-0.3cm}-4844.5 & \hspace{-0.3cm}-4492.0 & \hspace{-0.3cm}-4844.5 \\
    400	& 931.8 & 2795.2 & \hspace{-0.5cm}-4493.3 & \hspace{-0.3cm}-4845.9 & \hspace{-0.3cm}-4492.0 &  \hspace{-0.3cm}-4844.4 & \hspace{-0.3cm}-4492.0 & \hspace{-0.3cm}-4844.4 \\
    500	& 931.8 & 2795.2 & \hspace{-0.5cm}-4493.3 & \hspace{-0.3cm}-4845.9 & \hspace{-0.3cm}-4492.0 &  \hspace{-0.3cm}-4844.4 & \hspace{-0.3cm}-4492.0 & \hspace{-0.3cm}-4844.4 \\
    600	& 931.8 & 2795.2 & \hspace{-0.5cm}-4493.3 & \hspace{-0.3cm}-4845.9 & \hspace{-0.3cm}-4492.0 &  \hspace{-0.3cm}-4844.4 & \hspace{-0.3cm}-4492.0 &  \hspace{-0.3cm}-4844.4  \\
    700	& 931.8 & 2795.2 & \hspace{-0.5cm}-4493.3 & \hspace{-0.3cm}-4845.9 & \hspace{-0.3cm}-4492.0 &  \hspace{-0.3cm}-4844.4 & \hspace{-0.3cm}-4492.0 &  \hspace{-0.3cm}-4844.4  \\
    \hline\\[-0.35cm]
    $\nu-\nu^{[2]}$ $^\text{a}$  & -0.1 & -0.5 & 0.0 & -0.4 & 1.3 & 1.1 & 1.3 & 1.1  \\
    $\nu-\nu^{[4]}$ $^\text{b}$ &  &  &  &  & 0.1 &  -0.2 & 0.1 & -0.2  \\    
    \hline\hline \\[-0.35cm]
    $\nu_\mathrm{MCDF}$  \cite{boTheoreticalStudyInteresting2008} & 960 & 2900 & \hspace{-0.5cm}-4500 & \hspace{-0.3cm}-4920 & \hspace{-0.3cm}-4500 & \hspace{-0.3cm}-4920 & \hspace{-0.3cm}-4500 & \hspace{-0.3cm}-4920 \\
    \hline\hline
    \end{tabular}
    \begin{flushleft}
{\footnotesize    
    $^\text{a}$
      For the $\alpha^2\Eh$-order $\nu^{[2]}$, we used 
       $\nu_{10,\mathrm{DC}}^{[2]}
       =
       -(\langle H_\mathrm{so} \rangle + \langle H_\mathrm{soo} \rangle /3 )$, 
       $\nu_{20,\mathrm{DC}}^{[2]}
       =
       -3(\langle H_\mathrm{so} \rangle + \langle H_\mathrm{soo} \rangle/3)$,  
       $\nu_{10,\mathrm{DCB}}^{[2]}
       =
       -(\langle H_\mathrm{so} \rangle + \langle H_\mathrm{soo} \rangle)+ \langle H_\mathrm{ss} \rangle$, 
       and 
       $\nu_{20}^{\mathrm{DCB}[2]}=-3(\langle H_\mathrm{so} \rangle + \langle H_\mathrm{soo} \rangle)+ 9\langle H_\mathrm{ss} \rangle /5$ \cite{schwartzFineStructureHelium1964a,drakeProgressHeliumFinestructure2002}
       with
       $\langle H_\mathrm{so} \rangle =-0.034\ 659\ 207\ 422\ \alpha^2 \Eh$,
       $\langle H_\mathrm{soo} \rangle =0.051\ 478\ 075\ 59\ \alpha^2 \Eh$, and
       $\langle H_\mathrm{ss} \rangle =-0.022\ 520\ 165\ 86\ \alpha^2 \Eh$
       \cite{drakeProgressHeliumFinestructure2002}. \\
    $^\text{b}$       
      The $\nu^{[4]}_{J0}$ fine-structure splittings include all non-radiative contributions from the $^3P$-type basis states up to $\alpha^4\Eh$ order (including pair-correction and retardation effects). The $\alpha^4\Eh$-order correction was obtained as the sum of  the `subtotal contribution' of Table~4 and the $^3P$ contribution of Table~6 in Ref.~\cite{drakeProgressHeliumFinestructure2002}.
}    
    \end{flushleft}
  \end{table}

\begin{table}[ht]
    \caption{%
      Centroid energy, $E_\text{c}=(E_0+3E_1+5E_2)/9$ (with $E_J$ and $J=0,1,2$) in $\Eh$, for the $2~^3P^\texto_J$ manifold of the helium atom. 
    \label{tab:Hetripletcentroid}}
    \begin{tabular}{@{}l rrrr@{}}
         \hline \hline
         $N$ & 
         \multicolumn{1}{c}{$E_{\text{c},\DC}^\pp$} & 
         \multicolumn{1}{c}{$E_{\text{c},\DCpB}^\pp$} & 
         \multicolumn{1}{c}{$E_{\text{c},\DCppB}^\pp$} &  
         \multicolumn{1}{c}{$E_{\text{c},\DCB}^\pp$} \\ \hline
         300 & \num{-2.13327081123} & \num{-2.1332689432} & \num{-2.1332689434} & \num{-2.1332689434} \\
         400 & \num{-2.13327082648} & \num{-2.1332689584} & \num{-2.1332689587} & \num{-2.1332689587}\\
         500 & \num{-2.13327083251} & \num{-2.1332689645} & \num{-2.1332689647} & \num{-2.13326896471}\\
         600 & \num{-2.13327083299} & \num{-2.1332689649} & \num{-2.1332689652} & \num{-2.13326896519}\\ 
         700 & \num{-2.13327083309} & \num{-2.13326896504} & \num{-2.13326896529} & \num{-2.13326896529} \\ \hline 
         $E_\text{c}-E_\text{c}^{[2]}$ $^\text{a}$ & \num{-0.0000000097} & \num{-0.0000000097} & \num{-0.0000000100} & \num{-0.0000000100} \\
         $\varepsilon_\text{c}^{[4]}$ $^\text{b}$  &   &  &  \num{-0.0000000110}  & \num{-0.0000000110} \\
         \hline \hline
    \end{tabular}
    \begin{flushleft}
    \footnotesize{
      $^\text{a}$ %
        The $\alpha^2\Eh$-order centroid energy is  
        $E_\text{c,\text{DCB}}^{[2]}=-$2.133 268 955 3~$\Eh$~\cite{yerokhinAtomicStructureCalculations2021}. 
        The $\alpha^2\Eh$-order energy for the DC model, $E_\text{c,\text{DC}}^{[2]}=-$2.133 270 823 4~$\Eh$, was obtained from $E_\text{c,\text{DCB}}^{[2]}$ by subtracting the relevant terms. The orbit-orbit correction value was not reported (separately) in the literature, so we used $\langle H_\text{OO} \rangle=1.8681 \mu E_\text{h}$ evaluated in this work with the largest basis set.
        \\
      $^\text{b}$ %
        $\alpha^4\Eh$-order non-radiative correction for $^3P$-type basis states after cancelling divergences with some radiative corrections, $\varepsilon_\text{c}^{[4]}=E_Q+E_A'$ of Table~II in Ref.~\cite{pachuckiHeliumEnergyLevels2006} was used, which includes also pair-correction and retardation effects.
        }
    \end{flushleft}
\end{table}

\clearpage    
\subsubsection{\texorpdfstring{$^1S^\texte_0$}{} states of He, Li\texorpdfstring{$^+$}{}, and Be\texorpdfstring{$^{2+}$}{} \label{sec:singletatom}}
The ground state of light atoms with two electrons is primarily dominated by $^1S^\texte$-type (e: even parity) spatial functions and most importantly contributions from the $1~^1S^\texte$ non-relativistic wave function. %
This property motivates a relativistic description using the $LS$-coupling basis representation \cite{jeszenszkiAllorderExplicitlyCorrelated2021,jeszenszkiVariationalDiracCoulombExplicitly2022,ferencBreitInteractionExplicitly2022a,ferencVariationalVsPerturbative2022}.
To arrive at a complete relativistic basis space, in addition to the most important $^1 S^\texte$ ($S=0,L=0,p=+1$) functions, we must include also $^3P^\texte$-type ($S=1,L=1,p=+1$) functions in the spatial basis. (The $L=1,p=+1$ combination is sometimes called `unnatural parity', since the `natural parity' of a $Y_{LM}$ spherical harmonic function with $L=1$ is $-1$.)

In earlier work \cite{jeszenszkiAllorderExplicitlyCorrelated2021,jeszenszkiVariationalDiracCoulombExplicitly2022,ferencVariationalVsPerturbative2022}, we only estimated the triplet contributions (to be small) by adding (non-symmetry-adapted) basis functions to a large, ground-state ($J=0,p=+1$) relativistic computation by minimizing the DC$^{++}$ energy. This approach was not efficient enough to determine the precise contribution of the $^3P^\texte$ basis sector to the relativistic energy. 

In the present work, we compute the triplet contribution using symmetry-adapted, extensively optimized basis functions. 
Similarly to the $^1 S^\texte$ basis parameterization, the $^3P^\texte$ functions were parameterized by minimization of the non-relativistic energy of the lowest-energy state with this symmetry (convergence of the non-relativistic energy is shown in the \som).

$S^\texte$ and $P^\texte$-type basis functions were designed by a special parameterization of the fECGs and projection to the appropriate irreps of the $D_{2\text{h}}$ point group, a subgroup of $O(3)$ (following the subgroup-projection idea of Strasburger \cite{strasburgerHighAngularMomentum2014}),
which allowed us to use our already existing integral library~\cite{ferencNonadiabaticRelativisticLeadingOrder2020}.
Technical and implementation details are provided in the \som. 
In short, in this subgroup representation, the $S^\texte$-type spatial functions belong to the totally symmetric $A_\text{g}$ irrep, the $P^\texte_x$, $P^\texte_y$, and $P^\texte_z$ functions transform according to the $B_{3\text{g}}$, $B_{2\text{g}}$, and $B_{1\text{g}}$ irreps of $D_{2\text{h}}$ (which correspond to the irreps of rotation about the $x$, $y$, and $z$ axes), respectively.  

In the relativistic DC(B) computation, $J=0$ spin-spatial spinors are prepared by
combining the (symmetry-adapted) spatial functions with the (symmetry-adapted) spin functions, Eq.~(\ref{eq:Txyzdef}) (and Sec.~\ref{sec:doublegroup}),
resulting in the relevant basis states, $\varphi_{0,0}^{(1)} \Sigma_{0,0}$, $\varphi_{1,x}^{(1)}\Sigma_{1,x}$,  $\varphi_{1,y}^{(1)}\Sigma_{1,y}$, and $\varphi_{1,z}^{(1)}\Sigma_{1,z}$.

\begin{table}[ht]
  \centering
  \caption{%
    Triplet contribution, $\Delta E =E(N_\text{s},N_\text{t})-E(N_\text{s},0)$ in n$\Eh$, 
    to the $1\ ^1S^\texte_0$ energy of the helium atom
    with respect to the $N_\text{t}$ triplet and $N_\text{s}$ singlet basis set sizes. 
    Convergence of the total no-pair energy is currently limited by the singlet basis size, and corresponding convergence tables are shown in Table~S10 of the \som.
    \label{tab:hestdiff}
  }
  \begin{tabular}{@{}l@{\ \ } d{-1}d{-1}d{-1} c  d{-1}d{-1}d{-1} c  d{-1}d{-1}d{-1} c d{-1}d{-1}d{-1}@{}} 
    \hline \hline \\[-0.35cm]
     & 
     \multicolumn{3}{c} {$\Delta E_\mathrm{DC}^{++}$} & 
     & 
     \multicolumn{3}{c} {$\Delta E_{\DCpB}^{++}$} & 
     & 
     \multicolumn{3}{c} {$\Delta E_{\DCppB}^{++}$} & 
     & 
     \multicolumn{3}{c} {$\Delta E_\mathrm{DCB}^{++}$}  \\ 
     \cline{2-4}\cline{6-8}\cline{10-12}\cline{14-16} \\[-0.35cm]
    \hspace{0.2cm}$N_\text{t} \backslash N_\text{s}$ $^\text{a}$  & \multicolumn{1}{c} {400} & \multicolumn{1}{c}{500} & \multicolumn{1}{c}{700}  & \hspace{0.3cm} & \multicolumn{1}{c}{400} & \multicolumn{1}{c} {500} & \multicolumn{1}{c} {700} & \hspace{0.3cm} & \multicolumn{1}{c} {400} & \multicolumn{1}{c} {500} & \multicolumn{1}{c} {700} & \hspace{0.3cm} & \multicolumn{1}{c} {400} & \multicolumn{1}{c} {500} & \multicolumn{1}{c} {700}  \\ \hline
        \hspace{0.15cm}20 & -0.1 & -0.1 & -0.1 & & -0.1 & -0.1 & -0.1 & & -0.4 & -0.4 & -0.4  & & -0.4 & -0.4 & -0.4 \\
        \hspace{0.15cm}50 & -0.1 & -0.1 & -0.1 & & -0.1 & -0.1 & -0.1 & & -0.6 & -0.6 & -0.6 & & -0.6 & -0.6 & -0.6 \\
        \hspace{0.15cm}70 & -0.2 & -0.2 & -0.2 & & 0.0 & 0.0 & 0.0 & & -0.7 &  -0.7  &  -0.7 & & -0.7 &  -0.7  &  -0.7  \\
       100 & -0.2 & -0.2 & -0.2 & & 0.1 & 0.1 & 0.1 & &  -0.9  &  -0.9 &  -0.9 & &  -0.9  &  -0.9 &  -0.9\\
       150 & -0.2 & -0.2 & -0.2 & & 0.1 & 0.1 & 0.1 & & -0.9 & -0.9 & -1.0 &  & -0.9 & -0.9 & -1.0\\
       200 & -0.3 & -0.3 & -0.3 & & 0.1 & 0.1 & 0.1 & & -1.1 & -1.1 & -1.1 &  & -1.1 & -1.1 & -1.1\\
       250 & -0.3 & -0.3 & -0.3 & & 0.2 & 0.2 & 0.2 & & -1.1 & -1.1 & -1.1 & & -1.1 & -1.1 & -1.1\\ \hline
           $E_C $ \cite{pachuckiAlphaMathcalRCorrections2006} & \multicolumn{3}{c}{} & & \multicolumn{3}{c}{} & & \multicolumn{3}{c}{$-$1.113}  & & \multicolumn{3}{c}{$-$1.113} \\ \hline \hline
    \end{tabular}
    \begin{flushleft}
    {\footnotesize
      $^\text{a}$ %
        $N_\text{s}$ and $N_\text{t}$: number of fECGs used for the singlet and triplet  basis space, respectively. For the triplet states, $N_\text{t}$ functions are used for every ($x,y,z$) spatial direction. \\ 
      %
      %
      %
      %
      }
    \end{flushleft}
\end{table}

\begin{table}[ht]
  \centering
  \caption{%
    Triplet contribution, $\Delta E =E(N_\text{s},N_\text{t})-E(N_\text{s},0)$ in n$\Eh$, 
    to the $2\ ^1S^\texte_0$ energy of the helium atom
    with respect to the $N_\text{t}$ triplet and $N_\text{s}$ singlet basis set sizes. 
    Convergence of the total no-pair energy is currently limited by the singlet basis size and corresponding convergence tables are shown in Table~S11 of the \som.
    \label{tab:he2sstdiff}
  }
  \begin{tabular}{@{}l@{\ \ } ccc c  ccc c  ccc c  ccc@{}}
    \hline \hline \\[-0.35cm]
    & 
    \multicolumn{3}{c} {$\Delta E_\mathrm{DC}^{++}$} & \hspace{0.3cm} &
    \multicolumn{3}{c} {$\Delta E_{\DCpB}^{++}$} & 
    \hspace{0.3cm} & 
    \multicolumn{3}{c} {$\Delta E_{\DCppB}^{++}$} & 
    \hspace{0.3cm} &
    \multicolumn{3}{c} {$\Delta E_\mathrm{DCB}^{++}$} \\
    \cline{2-4}\cline{6-8}\cline{10-12}\cline{14-16} \\[-0.35cm]
    $N_\text{t} \backslash N_\text{s}$ $^\text{a}$ 
    & 400 & 500 &600  &  & 400 & 500 & 600 & & 400 & 500 & 600 & & 400 & 500 & 600  \\ \hline
        \hspace{0.15cm}20 & $-$0.01 & $-$0.01 & $-$0.01 & & $-$0.01 & $-$0.01 & $-$0.01 & & $-$0.03 & $-$0.03 & $-$0.02 & & $-$0.03 & $-$0.03 & $-$0.02 \\
        \hspace{0.15cm}50 & $-$0.01 & $-$0.01 & $-$0.01 & & \hspace{0.3cm}0.00 & \hspace{0.3cm}0.00 & \hspace{0.3cm}0.00 & & $-$0.05 & $-$0.05 & $-$0.05 & & $-$0.05 & $-$0.05 & $-$0.05  \\
        \hspace{0.15cm}70 & $-$0.01 & $-$0.01 & $-$0.01 & & \hspace{0.3cm}0.00 & \hspace{0.3cm}0.00 & \hspace{0.3cm}0.00 & & $-$0.06 &  $-$0.06  &  $-$0.06 & & $-$0.06 &  $-$0.06  &  $-$0.06 \\
       100 & $-$0.02 & $-$0.02 & $-$0.02 & & \hspace{0.3cm}0.01 & \hspace{0.3cm}0.01 & \hspace{0.3cm}0.01 & &  $-$0.08  &  $-$0.08 &  $-$0.07 & & $-$0.08  &  $-$0.08 &  $-$0.08 \\
       150 & $-$0.02 & $-$0.02 & $-$0.02 & & \hspace{0.3cm}0.01 & \hspace{0.3cm}0.01 & \hspace{0.3cm}0.01 & & $-$0.08 & $-$0.08 & $-$0.08 & & $-$0.08 & $-$0.08 & $-$0.08 \\
       200 & $-$0.02 & $-$0.02 & $-$0.02 & & \hspace{0.3cm}0.02 & \hspace{0.3cm}0.02 & \hspace{0.3cm}0.02 & & $-$0.09 & $-$0.09 & $-$0.09 & & $-$0.09 & $-$0.09 & $-$0.09  \\
       250 & $-$0.03 & $-$0.03 & $-$0.03 & & \hspace{0.3cm}0.03 & \hspace{0.3cm}0.03 & \hspace{0.3cm}0.03 & & $-$0.09 & $-$0.09 & $-$0.09 & & $-$0.09 & $-$0.09 & $-$0.09 \\ \hline
         $E_C $ \cite{pachuckiAlphaMathcalRCorrections2006} & \multicolumn{3}{c}{} & & \multicolumn{3}{c}{} & & \multicolumn{3}{c}{$-$0.09582}  & & \multicolumn{3}{c}{$-$0.09582} \\ 
      \hline \hline
    \end{tabular}
    \begin{flushleft}
    {\footnotesize
      $^\text{a}$
        $N_\text{s}$ and $N_\text{t}$: number of fECGs used for the singlet and triplet basis space, respectively. For the triplet states, $N_\text{t}$ functions are used for every ($x,y,z$) spatial direction. \\
%
    }
    \end{flushleft}
\end{table}

Tables~\ref{tab:hestdiff} and \ref{tab:he2sstdiff} show the convergence of the triplet energy contribution 
to the 1 and 2~$^1S^\texte_0$ energies of the helium atom
with respect to the number of the singlet and triplet basis functions.
We computed the `triplet contribution' (shown in the tables) as the difference of the variational no-pair energy obtained with the complete singlet and triplet basis space and the no-pair energy obtained using the singlet basis functions only. For the largest basis sets, the triplet contribution (difference) is converged to 0.1--0.01~n$\Eh$ (using double precision arithmetic). We also note that the 1 and 2~$^1S^\texte$ non-relativistic energies (\emph{cf.} Tables~S10-11 of the \som) are converged `only' to the 0.4~n$\Eh$ (with respect to more precise reference values available from the literature). The singlet basis functions used in the relativistic computation are parameterized based on the non-relativistic energy minimization, and thus, we think that by improving the non-relativistic basis parameterization, we could improve the convergence of the total relativistic energy as well.

Nevertheless, the triplet contribution appears to be sufficiently converged (Tables~\ref{tab:hestdiff} and \ref{tab:he2sstdiff}) and it is obtained in an excellent (numerical) agreement with the $\alpha^4\Eh$-order perturbative triplet correction~\cite{pachuckiAlphaMathcalRCorrections2006}.

Finally, the Breit contribution to the no-pair energy is dominated at the two-Breit photon level, which is shown by the fact that the ${\DCppB}^{++}$ and DCB$^{++}$ energies differ by less than 0.1~n$\Eh$, and thus, both results are in excellent agreement with the nrQED values, which account for the two Breit photon exchange in the $E_C$ correction \cite{pachuckiAlphaMathcalRCorrections2006} (cited in Tables~\ref{tab:hestdiff} and \ref{tab:he2sstdiff}).

\begin{table}[ht]
  \centering
  \caption{%
    Triplet contribution, $\Delta E =E(N_\text{s},N_\text{t})-E(N_\text{s},0)$ in n$\Eh$, 
    to the $1\ ^1S^\texte_0$ energy of the Li$^+$ ion
    with respect to the $N_\text{t}$ triplet and $N_\text{s}$ singlet basis set sizes. 
    Convergence of the total no-pair energy is currently limited by the singlet basis size and corresponding convergence tables are shown in Table~S12 of the \som.
    \label{tab:lipstdiff}
  }
  \begin{tabular}{@{}l@{\ \ } ccc c ccc c ccc c ccc@{}}
    \hline \hline \\[-0.35cm]
     & 
     \multicolumn{3}{c} {$\Delta E_\mathrm{DC}^{++}$} & 
     \hspace{0.5cm} & 
     \multicolumn{3}{c} {$\Delta E_{\DCpB}^{++}$} & 
     \hspace{0.5cm} & 
     \multicolumn{3}{c} {$\Delta E_{\DCppB}^{++}$} & 
     \hspace{0.5cm} & 
     \multicolumn{3}{c} {$\Delta E_\mathrm{DCB}^{++}$} \\ 
     \cline{2-4}\cline{6-8}\cline{10-12}\cline{14-16} \\[-0.35cm]
     $N_\text{t} \backslash N_\text{s}$ $^\text{a}$  
      & 400 & 500 &600  & & 400 & 500 & 600 & & 400 & 500 & 600 &&  400 & 500 & 600 \\ \hline
        \hspace{0.15cm}20 & $-$0.5 & $-$0.5 & $-$0.4 & & $-$0.6 & $-$0.7 & $-$0.6 & & $-$2.4 & $-$2.4 & $-$2.4 && $-$2.4 & $-$2.4 & $-$2.4 \\
        \hspace{0.15cm}50 & $-$0.8 & $-$0.8 & $-$0.8 & & $-$0.5 & $-$0.6 & $-$0.5 & & $-$4.1 & $-$4.1 & $-$4.1 && $-$4.1 & $-$4.0 & $-$4.1 \\
        \hspace{0.15cm}70 & $-$1.2 & $-$1.2 & $-$1.2 & & \hspace{0.3cm}0.1 & \hspace{0.3cm}0.1 & \hspace{0.3cm}0.2 & & $-$5.5 & $-$5.5 & $-$5.4 &&  $-$5.5 &  $-$5.5 &  $-$5.5 \\
       100 & $-$1.2 & $-$1.2 & $-$1.2 & & \hspace{0.3cm}0.1 & \hspace{0.3cm}0.1 & \hspace{0.3cm}0.1 & &  $-$5.5  &  $-$5.5 &  $-$5.5 &&  $-$5.5 &  $-$5.5 &  $-$5.5 \\
       150 & $-$1.6 & $-$1.6 & $-$1.6 & & \hspace{0.3cm}0.8 & \hspace{0.3cm}0.8 & \hspace{0.3cm}0.8  & & $-$6.5 & $-$6.5 & $-$6.5 &&  $-$6.5 & $-$6.5 & $-$6.5 \\
       200 & $-$2.0 & $-$2.0 & $-$1.9 & & \hspace{0.3cm}1.6 & \hspace{0.3cm}1.6 & \hspace{0.3cm}1.7  & & $-$7.3 & $-$7.4 & $-$7.3 && $-$7.3 & $-$7.3 & $-$7.3 \\ 
       250$^\text{f}$ & $-$2.1 & $-$2.1 & $-$2.1 & & \hspace{0.3cm}1.9 & \hspace{0.3cm}1.9 & \hspace{0.3cm}1.9 & & $-$7.6 & $-$7.6 & $-$7.6 && $-$7.6 & $-$7.6 & $-$7.6 \\
       300$^\text{f}$ & $-$2.2 &  $-$2.2 & $-$2.2 & & \hspace{0.3cm}2.1 & \hspace{0.3cm}2.1 &   \hspace{0.3cm}2.1 &  & $-$7.8 & $-$7.8 & $-$7.8 && $-$7.8 & $-$7.8 & $-$7.8\\ \hline \hline
    \end{tabular}
    \begin{flushleft}
    {\footnotesize
      $^\text{a}$ %
        $N_\text{s}$ and $N_\text{t}$: number of fECGs used for the singlet and triplet basis space, respectively. For the triplet states, $N_\text{t}$ functions are used for every ($x,y,z$) spatial direction. \\
      $^\text{b}$ %
        Quadruple precision had to be used due to near-linear dependencies in the basis set.
    }
    \end{flushleft}
\end{table}

\begin{table}[ht]
  \centering
  \caption{%
    Triplet contribution, $\Delta E =E(N_\text{s},N_\text{t})-E(N_\text{s},0)$ in n$\Eh$, 
    to the $1\ ^1S^\texte_0$ energy of the Be$^{2+}$ ion
    with respect to the $N_\text{t}$ triplet and $N_\text{s}$ singlet basis set sizes. 
    Convergence of the total no-pair energy is currently limited by the singlet basis size and corresponding convergence tables are shown in Table~S13 of the \som.    
     \label{tab:be2pstdiff}
  }
  \begin{tabular}{@{}l@{\ \ } ccc c  ccc c  ccc c  ccc @{}}
     \hline \hline \\[-0.35cm]
     & 
     \multicolumn{3}{c} {$\Delta E_\mathrm{DC}^{++}$} & 
     \hspace{0.5cm} & 
     \multicolumn{3}{c} {$\Delta E_{\DCpB}^{++}$} & 
     \hspace{0.5cm} & 
     \multicolumn{3}{c} {$\Delta E_{\DCppB}^{++}$} & 
     \hspace{0.5cm} & 
     \multicolumn{3}{c} {$\Delta E_\mathrm{DCB}^{++}$} \\ 
     \cline{2-4}\cline{6-8}\cline{10-12}\cline{14-16} \\[-0.35cm]
      $N_\text{t} \backslash N_\text{s}$ $^\text{a}$ & 
      100 & 200 &300  &  & 100 & 200 & 300 & & 100 & 200 & 300 & & 100 & 200 & 300  \\ \hline
        \hspace{0.15cm}20 & $-$2.1 & $-$2.0 & $-$2.1 & & $-$0.1 & $-$0.1 & $-$0.1 & & $-$10 & $-$10 & $-$10 & & $-$10 & $-$10 & $-$10 \\
        \hspace{0.15cm}50 & $-$3.8 & $-$3.7 & $-$3.9 & & $-$0.1 & $-$0.1 & $-$0.1 & & $-$18 & $-$18 & $-$18 & & $-$18 & $-$18 & $-$17  \\
        \hspace{0.15cm}70 & $-$3.9 & $-$3.8 & $-$4.2 & & $-$0.5 & $-$0.4 & $-$0.7 & & $-$18 &  $-$18  &  $-$18 & & $-$18 &  $-$18  &  $-$18 \\
       100 & $-$4.5 & $-$4.4 & $-$4.6 & & \hspace{0.3cm}1.0 & \hspace{0.3cm}1.0 & \hspace{0.3cm}0.8 & &  $-$19  &  $-$19 &  $-$19 & & $-$19  &  $-$19 &  $-$19 \\
       150 & $-$6.3 & $-$6.2 & $-$6.3 & & \hspace{0.3cm}3.9 & \hspace{0.3cm}4.0 & \hspace{0.3cm}3.9 & & $-$25 & $-$25 & $-$25 & & $-$24 & $-$25 & $-$24 \\
       200 & $-$6.2 & $-$6.2 & $-$6.3 & & \hspace{0.3cm}3.9 & \hspace{0.3cm}3.9 & \hspace{0.3cm}3.8 & & $-$24 & $-$24 & $-$25 & & $-$24 & $-$24 & $-$24  \\
       250 & $-$6.4 & $-$6.2 & $-$6.6 & & \hspace{0.3cm}3.9 & \hspace{0.3cm}4.0 & \hspace{0.3cm}3.7 & & $-$25 & $-$25 & $-$25 & & $-$25 & $-$25 & $-$25 \\
      \hline \hline
  \end{tabular}
  \begin{flushleft}
    {\footnotesize
      $^\text{a}$ %
        $N_\text{s}$ and $N_\text{t}$: number of fECGs used for the singlet and triplet basis space, respectively. For the triplet states, $N_\text{t}$ functions are used for every ($x,y,z$) spatial direction.  \\
    }
  \end{flushleft}    
\end{table}

Tables~\ref{tab:lipstdiff} and \ref{tab:be2pstdiff} show the convergence of the triplet contribution to the ground-state energy of the Li$^+$ and Be$^{2+}$ ions, respectively (see also Tables~S12 and S13). For these systems, the complete $\alpha^4\Eh$-order perturbative correction has been reported \cite{yerokhinTheoreticalEnergiesLowlying2010,yerokhinQEDCalculationsEnergy2022a}, but the contributions that would be relevant for the present work are not separately available.

Along the $Z=2,3,4$ series of nuclear charge numbers, the relative importance of the triplet basis sector in the total ground-state energy increases from 0.4~ppb to 1.7~ppb. Further work is in progress regarding the intermediate and higher $Z$ range.

\begin{figure}
    \centering
    \includegraphics[scale=0.7]{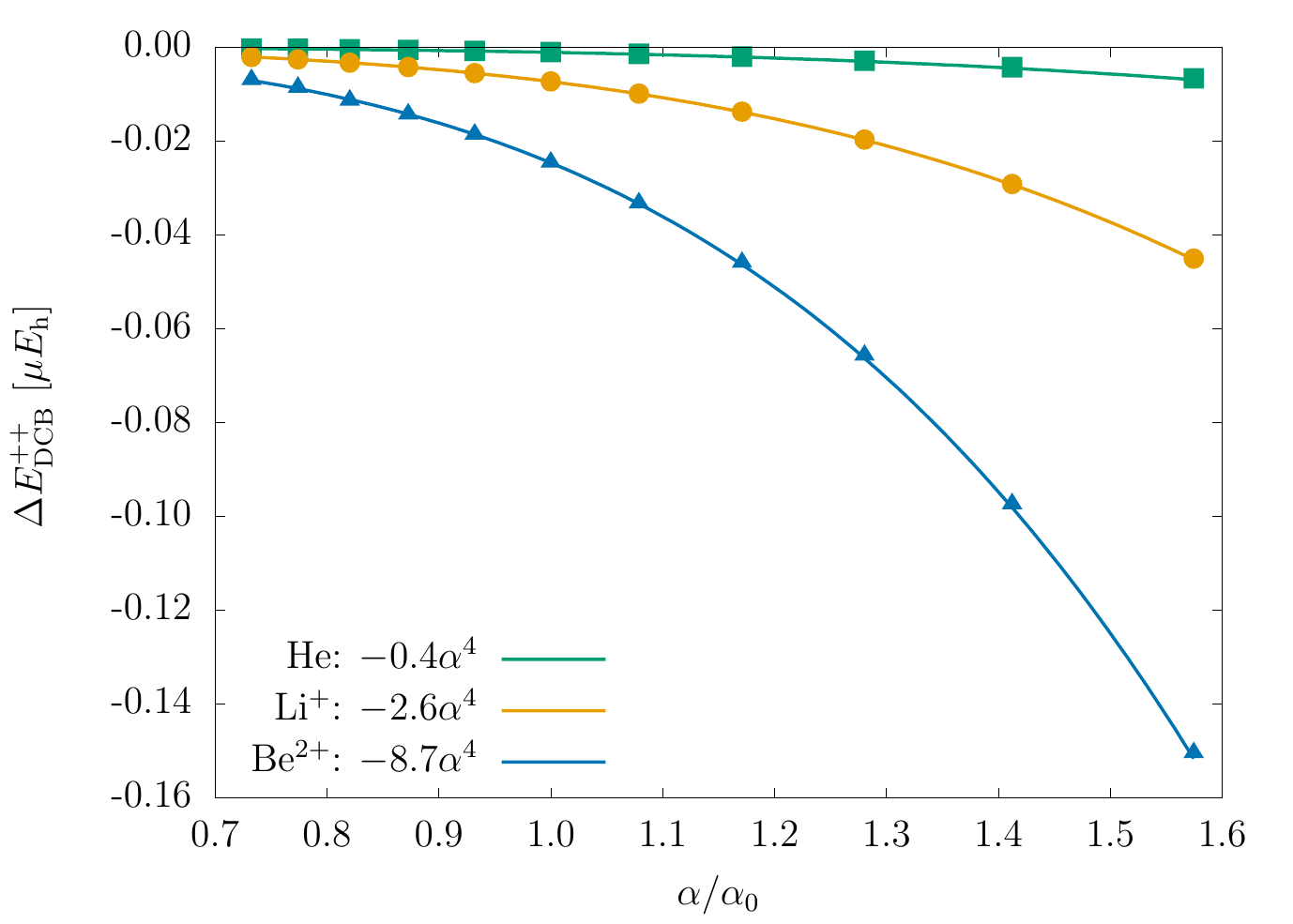}
    \caption{%
    Fine-structure constant dependence of the triplet energy contribution, $\Delta E^\pp_\mathrm{DCB}$, to the no-pair DCB energy. 
    The data points follow an $\alpha^4$ polynomial and the fitted prefactor reproduces the nrQED result available for the ground-state helium atom, $E_C=-0.392\ 621 \alpha^4 \Eh \approx-0.4 \alpha^4 \Eh$~\cite{pachuckiAlphaMathcalRCorrections2006}.  
    The no-pair DCB equation was solved for $\alpha=1/(\alpha_0+10n)$, $n=-5, \dots, 5$ with $\alpha_0^{-1}=137.$035 999 084 \cite{Codata2018Recommended}
    (with $N_\text{s}=400$ for He and Li$^+$, $N_\text{s}=300$ for Be$^{2+}$ and $N_\text{t}=200$ for every spatial direction). For the He and Be$^{2+}$ computations double precision was sufficient, but we had to use 10-byte reals (20-byte complex numbers) for the Li$^+$ parameterization. 
    } 
  \label{fig:DCBPTrelsym}
\end{figure}

We have also checked the fine-structure constant ($\alpha$) dependence of the triplet energy contributions by repeating the no-pair variational computations with slightly different $\alpha$ values (Fig.~\ref{fig:DCBPTrelsym}). The triplet contribution is well described by a $t\alpha^4$ function and the fitted $t$ coefficient for the helium atom is in excellent agreement the formal perturbation theory result \cite{pachuckiAlphaMathcalRCorrections2006}.

Since, in this work, we use the complete relativistic basis space, we have considered again (cf.~Ref.~\cite{jeszenszkiVariationalDiracCoulombExplicitly2022}) comparison with the no-pair Dirac--Coulomb computations in a Hylleraas basis \cite{bylickiRelativisticHylleraasConfigurationinteraction2008}.  
For all three $Z=$ 2, 3, and 4 values, an 80~ppb relative energy difference persists between our DC$^{++}$ energies and those of Ref.~\cite{bylickiRelativisticHylleraasConfigurationinteraction2008}. 
A possible source of deviation may be due to the limited precision of the iterative kinetic balance condition and its relation to the Lévy--Leblond equation in the non-relativistic limit of Ref.~\cite{pestkaHylleraasCIApproachDiraccoulomb2003,pestkaApplicationComplexcoordinateRotation2006}. Despite the persisting (small, but non-negligible) disagreement with this other set of variational-type computations, the excellent agreement of the $\alpha$ dependence of our results (Refs.~\cite{jeszenszkiVariationalDiracCoulombExplicitly2022,ferencVariationalVsPerturbative2022} and this work) with the perturbative values provides a solid benchmark for our developments.

\begin{table}[ht]
  \centering
  \caption{%
    Triplet contribution, $\Delta E =E(N_\text{s},N_\text{t})-E(N_\text{s},0)$ in 0.1 n$\Eh$, 
    to the $X\ ^1\Sigma_\text{g}^+$ ground state energy of the H$_2$ molecule with $R_\text{pp}=1.4$~bohr nuclear separation with respect to the $N_\text{t}$ triplet and $N_\text{s}$ singlet basis set sizes. 
    Convergence of the total no-pair energy is currently limited by the singlet basis size, and corresponding convergence tables are shown in Table~S14 of the \som.
    \label{tab:h2stdiff}
  }
\scalebox{0.93}{%
  \begin{tabular}{@{}l l@{} c ccc c ccc c  ccc c ccc@{}}
    \hline\hline \\[-0.35cm]
     & & & 
     \multicolumn{3}{c}{$\Delta E_\mathrm{DC}^\pp$} & 
     \hspace{0.15cm} & 
     \multicolumn{3}{c}{$\Delta E_{\DCpB}^\pp$} & 
     \hspace{0.15cm} & 
     \multicolumn{3}{c} {$\Delta E_{\DCppB}^\pp$} & 
     \hspace{0.15cm} & 
     \multicolumn{3}{c}{$\Delta E_\mathrm{DCB}^\pp$}
     \\ 
     \cline{4-6}\cline{8-10}\cline{12-14}\cline{16-18} \\[-0.35cm]
     $N_{\Pi_\text{g}}$$^\text{a}$ &  ${N_{\Sigma_\text{g}^-}}^\text{a}\backslash N_{\Sigma_\text{g}^+}$$^\text{a}$: &  & 800 & 1000 &1200 & & 800 & 1000 & 1200 & & 800 & 1000 & 1200 & & 800 & 1000 & 1200 \\ \hline\\[-0.35cm]
     \multicolumn{8}{l}{$^3\Pi_\text{g}$ contributions:}\\
       \hspace{0.15cm}200 & & & $-$0.18 & $-$0.18 & $-$0.17 & & \hspace{0.3cm}0.00 & \hspace{0.3cm}0.00 & $-$0.01 & & $-$0.60 & $-$0.59 & $-$0.59 &  & $-$0.60 & $-$0.59 & $-$0.59 \\
       \hspace{0.15cm}300 & & & $-$0.22 & $-$0.22 & $-$0.21 & & \hspace{0.3cm}0.00 & \hspace{0.3cm}0.00 & \hspace{0.3cm}0.00 & & $-$0.75 & $-$0.75 & $-$0.74 & & $-$0.76 & $-$0.74 & $-$0.75  \\
       \hspace{0.15cm}400 & & & $-$0.24 & $-$0.24 & $-$0.23 & & \hspace{0.3cm}0.04 & \hspace{0.3cm}0.04 & \hspace{0.3cm}0.05 & & $-$0.74 & $-$0.74 & $-$0.72 & & $-$0.74 & $-$0.73 & $-$0.72  \\
       \hspace{0.15cm}600 & & & $-$0.29 & $-$0.28 & $-$0.26 &&  \hspace{0.3cm}0.10 & \hspace{0.3cm}0.11 & \hspace{0.3cm}0.13 &&  $-$0.85 & $-$0.84 & $-$0.83 & & $-$0.86 & $-$0.84 & $-$0.83 \\
       \hspace{0.15cm}800 & & & $-$0.33 & $-$0.34 & $-$0.33 & & \hspace{0.3cm}0.25 & \hspace{0.3cm}0.24 & \hspace{0.3cm}0.26 & & $-$0.96 & $-$0.97 & $-$0.96 & & $-$0.97 & $-$0.97 & $-$0.96 \\
       1000 & & & $-$0.34 & $-$0.34 & $-$0.34 & & \hspace{0.3cm}0.25 &  \hspace{0.3cm}0.25 &  \hspace{0.3cm}0.25 & & $-$0.96 & $-$0.96 & $-$0.96 & & $-$0.97 & $-$0.96 & $-$0.95  \\
       1200 & & & $-$0.35 &  $-$0.35 &  $-$0.33 & & \hspace{0.3cm}0.26 & \hspace{0.3cm}0.26 & \hspace{0.3cm}0.28 & & $-$0.99 &  $-$0.99 &  $-$0.97 & & $-$1.00 &  $-$0.98 &  $-$0.98 \\
    \hline\\[-0.35cm] 
    \multicolumn{8}{l}{$^3\Sigma_\text{g}^-$ contributions:}\\    
       & \hspace{0.15cm}80 & & $-$0.11 & $-$0.11 & $-$0.10 & & \hspace{0.3cm}0.03 & \hspace{0.3cm}0.03 & \hspace{0.3cm}0.03 & & $-$0.43 & $-$0.43 & $-$0.43 & & $-$0.42 & $-$0.42 & $-$0.42  \\
       & 100 & & $-$0.11 & $-$0.11 & $-$0.10 & & \hspace{0.3cm}0.02 & \hspace{0.3cm}0.02 & \hspace{0.3cm}0.03 & & $-$0.43 & $-$0.44 & $-$0.43 & & $-$0.43 & $-$0.43 & $-$0.43  \\
       & 140 & & $-$0.14 & $-$0.14 & $-$0.14 & & \hspace{0.3cm}0.10 & \hspace{0.3cm}0.09 & \hspace{0.3cm}0.10 & & $-$0.50 & $-$0.50 & $-$0.49 & & $-$0.49 & $-$0.49 & $-$0.48  \\
       & 200 & &$-$0.13 & $-$0.15 & $-$0.13 & & \hspace{0.3cm}0.10 & \hspace{0.3cm}0.08 & \hspace{0.3cm}0.10 & & $-$0.49 & $-$0.50 & $-$0.48 & & $-$0.50 & $-$0.52 & $-$0.49  \\
       & 300 & & $-$0.17 & $-$0.16 & $-$0.15 &&  \hspace{0.3cm}0.14 & \hspace{0.3cm}0.15 & \hspace{0.3cm}0.16 &&  $-$0.53 & $-$0.52 & $-$0.51 & & $-$0.52 & $-$0.52 & $-$0.52 \\
       & 400 & & $-$0.16 & $-$0.16 & $-$0.15 & & \hspace{0.3cm}0.15 & \hspace{0.3cm}0.15 & \hspace{0.3cm}0.16 & & $-$0.53 & $-$0.53 & $-$0.52 & & $-$0.53 & $-$0.52 & $-$0.51 \\
    \hline\\[-0.35cm]
    \multicolumn{8}{l}{$^3\Pi_\text{g}$ and $\Sigma_\text{g}^-$ contributions:}\\        
        \hspace{0.15cm}800  & 300 & & $-$0.49 & $-$0.50 & $-$0.49 & & \hspace{0.3cm}0.41 & \hspace{0.3cm}0.40 & \hspace{0.3cm}0.41 & & $-$1.48 & $-$1.49 & $-$1.48 & & $-$1.49 & $-$1.48 & $-$1.48 \\
        1000  & 300 & & $-$0.50 & $-$0.50 & $-$0.49 & & \hspace{0.3cm}0.40 & \hspace{0.3cm}0.40 & \hspace{0.3cm}0.40 & & $-$1.48 & $-$1.48 & $-$1.48 & & $-$1.48 & $-$1.47 & $-$1.49 \\
        1000  & 400 & & $-$0.50 & $-$0.50 & $-$0.48 & & \hspace{0.3cm}0.39 & \hspace{0.3cm}0.39 & \hspace{0.3cm}0.41 & & $-$1.49 & $-$1.49 & $-$1.47 & & $-$1.50 & $-$1.49 & $-$1.48 \\
       1200  & 300 & & $-$0.51 & $-$0.50 & $-$0.49 & & \hspace{0.3cm}0.41 & \hspace{0.3cm}0.42 & \hspace{0.3cm}0.43 & & $-$1.51 & $-$1.51 & $-$1.49 & & $-$1.51 & $-$1.50 & $-$1.51 \\
       1200  & 400 & & $-$0.50 & $-$0.50 & $-$0.50 & & \hspace{0.3cm}0.41 & \hspace{0.3cm}0.41 & \hspace{0.3cm}0.41 & & $-$1.51 & $-$1.51 & $-$1.51 & & $-$1.52 & $-$1.51 & $-$1.51 \\
       \hline\\[-0.35cm] 
           \multicolumn{2}{l}{$E_C $ \cite{puchalskiCompleteAlpha6m2016}}  & & \multicolumn{3}{c}{} & & \multicolumn{3}{c}{} & & \multicolumn{3}{c}{}  & & \multicolumn{3}{c}{$-$1.69} \\
       \hline \hline
  \end{tabular}
}
  \begin{flushleft}
    {\footnotesize
      $^\text{a}$ %
        $N_{\Sigma_\text{g}^+}$, $N_{\Pi_\text{g}}$, and $N_{\Sigma_\text{g}^-}$ label the number of $\Sigma_\text{g}^+$-,  $\Pi_\text{g}$- and $\Sigma_\text{g}^-$-type spatial functions, respectively. For $\Pi_\text{g}$-type functions, $N_{\Pi_\text{g}}$ functions are used for both ($x,y$) spatial directions.
    }
  \end{flushleft}        
\end{table}

\subsection{Computations for the ground state of the H\texorpdfstring{$_2$ \label{sec:h2molecule}}{}  molecule}
In earlier work \cite{jeszenszkiAllorderExplicitlyCorrelated2021,jeszenszkiVariationalDiracCoulombExplicitly2022,ferencBreitInteractionExplicitly2022a,ferencVariationalVsPerturbative2022}, we reported well-converged DC(B) energies using the $\Sigma_\text{g}^+ \Sigma_{0,0}$ basis sector, which gives the dominant contribution to the ground-state energy of H$_2$. According to formal perturbation theory \cite{puchalskiCompleteAlpha6m2016}, the triplet basis sector contribution starts at $\alpha^4\Eh$ order.

In the present work, we determine the triplet contribution to the ground-state energy of H$_2$ by using the entire relativistic basis space in the variational computation. Similarly to the atomic computations (Sec.~\ref{sec:heliumlike}), this can be efficiently realized by exploiting the (double-group) symmetry properties of the system.

The relativistic states of H$_2$, a homonuclear diatomic molecule transform according to the $D_{\infty \text{h}}$ double-point group. 
The complete relativistic basis space is spanned by 
the (dominant) $\Sigma_\text{g}^+\Sigma_{0,0}$-type functions
and by 
$\Pi_{\text{g},x}\Sigma_{1,x}$, $\Pi_{\text{g},y}\Sigma_{1,y}$, 
and $\Sigma_\text{g}^- \Sigma_{1,z}$-type spatial-spin functions.

For the numerical, no-pair DC(B) computations presented in this work, we used (a special parameterization of) fECG functions and sub-group projection (introduced for atoms by Strasburger \cite{strasburgerHighAngularMomentum2014}). Similarly to the atomic computations (Sec.~\ref{sec:heliumlike}), we used the $D_{\text{2h}}$ group as a subgroup of $D_{\infty\text{h}}$ to construct symmetry-adapted basis functions for H$_2$. 
This special construction (for further details see also the \som) allowed us to use the elementary integrals already available in QUANTEN~\cite{ferencNonadiabaticRelativisticLeadingOrder2020}.

In the $D_{\text{2h}}$ group, the $\Sigma_\text{g}^+$-type functions of $D_{\infty\text{h}}$ are represented according to the $A_g$ totally symmetric irrep. 
As to the spatial functions relevant for triplet contributions to the relativistic ground state, 
the $\Sigma_\text{g}^-$-type functions are represented according to the $B_{1\text{g}}$ irrep, while 
the $\Pi_{\text{g},x}$ and $\Pi_{\text{g},y}$ transform according to the $B_{3\text{g}}$ and  $B_{2\text{g}}$ irreps in $D_{\text{2h}}$, respectively. 
Interestingly, rotations about the $z, x,$ and y axes transform according to these irreps,
$B_{1\text{g}}$, $B_{3\text{g}}$, and $B_{2\text{g}}$, respectively (Table~S1).

Then, a DC(B) computation with the complete relativistic basis space was carried out by including all basis types, and the convergence of the total DC(B) energy was checked with respect to the size of the three different types of the spatial basis sets, $\Sigma_\text{g}^+$, $\Pi_\text{g}$, and $\Sigma_\text{g}^-$.  The spatial basis functions were parameterized
by minimization of the lowest non-relativistic energy in the corresponding spatial irrep and spin state (Table~S5 and S6).

Table~\ref{tab:h2stdiff} shows the convergence of the relativistic energy.
It is interesting to observe that the triplet contributions to the ground-state energy due to the $\Pi_{\text{g}}$-type and the $\Sigma_\text{g}^-$-type spatial functions is almost additive (cf. the upper two `blocks' of the table and the lowest `block').
This observation is in agreement with the formal perturbation theory result, in which the triplet correction is obtained at leading $\alpha^4\Eh$-order as separate contributions due to  the triplet basis sectors of different symmetry (without any account for the coupling of the different triplet basis spaces)~\cite{drakeProgressHeliumFinestructure2002,yerokhinAtomicStructureCalculations2021,puchalskiCompleteAlpha6m2016}. 

We obtain the energy contribution to $\Delta E_{\DCppB}^{++}$ and $\Delta E_\DCB^{++}$ due to the coupling of $\Sigma_\text{g}^+\Sigma_{0,0}$ with the $\Pi_{\text{g},x} \Sigma_{1,x}$ and $\Pi_{\text{g},x} \Sigma_{1,y}$ basis states to be 0.1 n$\Eh$, and it is 0.05 n$\Eh$ due to the coupling of the $\Sigma_\text{g}^+\Sigma_{0,0}$ with the $\Sigma_\text{g}^-\Sigma_{1,z}$ basis sector.
In our variational computations, coupling of the different basis sectors is automatically included and it is found to be very small (less than 10~p$\Eh$) for this system. In all these variational computations, use of double precision arithmetic was found to be sufficient for the reported precision.

All in all, our no-pair ${\DCpB}$ and DCB computations result in a 0.15~n$\Eh$ 
triplet contribution to the ground-state energy of H$_2$, which is in reasonable agreement with the perturbative result of 0.169~n$\Eh$ \cite{puchalskiCompleteAlpha6m2016}. 
To address the remaining small deviation between the variational and formal perturbative numerical results, it would be necessary to further improve our basis representation, most importantly by more tightly optimizing the non-relativistic energy (Table~S14).

\clearpage
\section{Conclusion and outlook}
\noindent%
The no-pair Dirac--Coulomb--Breit wave equation is solved for two-electron, light atoms and molecules using the $LS$-coupling and double-group symmetry, which makes it possible
to efficiently parameterize the employed explicitly correlated Gaussian basis set based on non-relativistic energy optimization.
This paper completes our previous work \cite{jeszenszkiAllorderExplicitlyCorrelated2021,jeszenszkiVariationalDiracCoulombExplicitly2022,ferencBreitInteractionExplicitly2022a,ferencVariationalVsPerturbative2022}, in which only the dominant singlet basis sector was included in the relativistic basis set.
The no-pair Dirac--Coulomb and Dirac--Coulomb--Breit energies of helium (1 and $2\ S^\texte_0$ states), 
Li$^+$ and Be$^{2+}$ ($1\ S^\texte_0$ states), and the ground state of the H$_2$ molecule with fixed protons are converged within a sub-parts-per-billion relative precision including the complete relativistic (singlet and triplet) basis space. 
%
The $\alpha$ fine-structure constant dependence of the triplet contributions to the dominantly singlet relativistic energies is numerically determined from the no-pair computations to start at $\alpha^4\Eh$ order in excellent agreement, also in terms of the numerical values, with the formal perturbation theory results \cite{pachuckiAlphaMathcalRCorrections2006,puchalskiCompleteAlpha6m2016}. 
Further extension of the computations to medium (and high) $Z$ nuclear charge numbers of the helium isoelectronic series is planned for future work.

\vspace{0.5cm}
\section*{Data availability statement}
\noindent%
The data that support findings of this study is included in the paper or in the Supplementary Material.

\vspace{0.5cm}
\section*{Supplementary Material}
\noindent%
The \som\ contains 
(a) implementation details of the projection of floating ECG functions to irreps of finite subgroups of $O(3)$ and $D_{\infty\text{h}}$; (b) convergence tables.

\vspace{0.5cm}
\begin{acknowledgments}
\noindent Financial support of the European Research Council through a Starting Grant (No.~851421) is gratefully acknowledged. 
\end{acknowledgments}

\clearpage

\setcounter{section}{0}
\renewcommand{\thesection}{S\arabic{section}}
\setcounter{subsection}{0}
\renewcommand{\thesubsection}{S\arabic{section}.\arabic{subsection}}

\setcounter{equation}{0}
\renewcommand{\theequation}{S\arabic{equation}}

\setcounter{table}{0}
\renewcommand{\thetable}{S\arabic{table}}

\setcounter{figure}{0}
\renewcommand{\thefigure}{S\arabic{figure}}

~\\[0.cm]
\begin{center}
\begin{minipage}{0.8\linewidth}
\centering
\textbf{Supplementary Material} \\[0.25cm]

\textbf{%
Relativistic two-electron atomic and molecular energies using $LS$
coupling and double groups: role of the triplet contributions to singlet
states}
\end{minipage}
~\\[0.5cm]
\begin{minipage}{0.6\linewidth}
\centering

P\'eter Jeszenszki$^1,\ast$ and Edit M\'atyus$^{1,\dagger}$ \\[0.15cm]

$^1$~\emph{ELTE, Eötvös Loránd University, Institute of Chemistry, 
Pázmány Péter sétány 1/A, Budapest, H-1117, Hungary} \\[0.15cm]
$^\ast$ peter.jeszenszki@ttk.elte.hu \\
$^\dagger$ edit.matyus@ttk.elte.hu \\
\end{minipage}
~\\[0.15cm]
\end{center}

~\\[1cm]
\begin{center}
\begin{minipage}{0.9\linewidth}
\noindent %
Contents: \\
S1. Projection of floating {ECG} functions to irreps of finite subgroups of $O(3)$ and $D_{\infty\text{h}}$ \\
S2. Convergence tables 
\end{minipage}
\end{center}

\clearpage

\section{Projection of floating ECG functions to irreps of finite subgroups of \texorpdfstring{$O(3)$}{} and \texorpdfstring{$D_{\infty \mathrm{h}}$}{} \label{sec:symmetryadapt}}
\noindent
In this section the implementation of projectors to the $S^\texte$, $P^\texto$, and (unnatural-parity) $P^\texte$ irreps of the $O\mathrm{(3)}$ point group as well as to the $\Sigma_\text{g}^+$, $\Pi_\text{g}$, and $\Sigma_\text{g}^-$ irreps of the $D_{\infty \text{h}}$ point group is presented. 

The $O(3)$ group is the relevant symmetry group for atoms (and also the pre-Born--Oppenheimer problem). 
For a description of the various symmetry species, one could use a spherically symmetric ECG multiplied with some $\rho_n$ radial prefactor and an appropriate $\eta_\mathrm{sph}^{(\Gamma)}$ angular prefactor,
\begin{align}
  f_\mathrm{sph}^{(\Gamma)}(\bos{A},n)
  =
  \eta_\mathrm{sph}^{(\Gamma)} \rho_n 
  \exp\left[-\bos{r}^\text{T}\underline{\bos{A}} \bos{r}\right]\; .
  \label{eq:polangpref}
\end{align}
The $\rho_n$ term is essential for an efficient representation of nodes or particle densities centered off the origin, and $\eta^{(\Gamma)}$ 
(with $\Gamma$ collecting the spatial quantum numbers)
carries the spatial symmetry properties  \cite{vargaGlobalVectorRepresentationAngular1998,suzukiStochasticVariationalApproach1998,matyusMolecularStructureCalculations2012a,joyceMatrixElementsExplicitly2016,matyusPreBornOppenheimerMolecular2019}.

The $D_{\infty \mathrm{h}}$ point group is relevant in the case of homonuclear diatomics of cylindrical symmetry.
If the nuclei are fixed along the $z$ axis, displaced symmetrically about the origin,
the symmetry-adapted basis function can be written as
\begin{align}
  f_\mathrm{cyl}^{(\Omega)}(\bos{A},n)
  =
  \eta_\mathrm{cyl}^{(\Omega)} \xi_n(\bos{q},\bos{z}) 
  \exp\left[-\left(\bos{r}^\text{T}-\bos{s}^{(z)\text{T}}\right)\underline{\bos{A}} \left(\bos{r}-\bos{s}^{(z)}\right)\right] \; ,
  \label{eq:polangpcylref}
\end{align}
where $\eta_\mathrm{cyl}^{(\Omega)}$ describes the spherical part in two dimensions with the quantum numbers collected in $\Omega$. 
The vector $\bos{s}^{(z)}$ shifts the center of the ECG along the $z$ axis.
$\xi_n(\bos{q},\bos{z})$ describes the behavior of the wave function  perpendicular to the $z$ axis ($q_i=\sqrt{x_i^2+y_i^2}$ for particle $i$) for a given $z$ coordinate.
At the moment, we do not have an integral library for a general (any $\Gamma$ or $\Omega$) basis set for the relativistic matrix elements, so we considered an alternative procedure.

Efficient calculations for $O$(3) can be performed with analytic matrix elements using Cartesian prefactors in front of an ECG,
\begin{align}
  g(\bos{A},n,m,o)
  =
  x_i^n y_j^m z_k^o
  \exp\left[-\bos{r}^\text{T}\underline{\bos{A}}\bos{r}\right] \; ,
  \label{eq:cartpref}  
\end{align}
corresponding to some well-defined and not too high $n,m,o$ integers 
\cite{sharkeyAnalyticalEnergyGradient2010,sharkeyAlgorithmQuantumMechanical2013,puchalskiExplicitlyCorrelatedWave2015,puchalskiRelativisticCorrectionsGround2017}, and then, use linear combination of these functions which transform according to the desired irrep.

Another alternative option is provided by the floating ECGs, which have an appealing simple form
\begin{align}
  \Theta(\bos{A},\bos{s})
  =
  \exp\left[-(\bos{r}-\bos{s})^\text{T}\underline{\bos{A}} (\bos{r}-\bos{s})\right]\; ,
\end{align}
for which analytic integrals can be implemented for most physically relevant operators and $\bos{s}\neq0$ allows to efficiently describe particle densities off centered from the origin. This property can be used to replace high polynomial powers in Eq.~(\ref{eq:polangpref}). 
Unfortunately, fECGs are not eigenfunctions of the relevant symmetry groups (with $\bos{s}\neq 0$). For efficient computations, it is necessary to adapt them to the relevant irreps. 
Direct numerical projection to $O(3)$ irreps has been implemented with numerical integration to the three rotational angles \cite{muoloExplicitlyCorrelatedGaussian2018a}.

In this work, as a computationally less demanding and technically simple procedure, we follow Strasburger's finite subgroup projection approach developed for atoms \cite{strasburgerHighAngularMomentum2014,strasburgerExplicitlyCorrelatedWave2019}. 
Then, for computations to H$_2$ with fixed protons, we adapt his finite subgroup approach to the $D_{\infty\text{h}}$ point group (Sec.~\ref{sec:dinfth}). 

In our implementation, it is exploited that symmetry transformation of an fECG function, amounts to transforming its $\bos{s}$ vectors,
\begin{align}
  O \Theta_i(\br)
  &=
  \exp\left[%
    -\left(O^{-1}\br- \bs_i\right)^\tT 
    \underline{\bA}_i
    \left(O^{-1}\br-\bs_i\right)
  \right] 
  \nonumber \\
  &=
  \exp\left[%
    -\left(\br- \underbrace{O\bs_i}_{\bs_i^{O}} \right)^\tT 
    \underbrace{O^{-\tT} \underline{\bA}_i O^{-1}}_{\underline{\bA}_i} 
    \left(\br-\underbrace{O \bs_i}_{\bs_i^{O}} \right)
  \right]  
  \nonumber \\
  &=
  \exp\left[%
    -\left(\br- \bs_i^{O} \right)^\tT
    \underline{\bA}_i
    \left(\br-\bs_i^{O} \right)
  \right] \ ,
  \label{eq:spatialECGtransform}
\end{align}
where $\underline{\bos{A}}_i=\bos{A}_i\otimes \bos{1}_3$. 
(For deformed ECGs \cite{beutelDeformedExplicitlyCorrelated2021b}, this property does not hold and the deformed $\bos{A}$ matrix must be transformed as well. This transformation is certainly feasible and would only require some additional matrix operations.)
Hence, evaluation of the finite subgroup projectors requires only evaluation of a finite sum of fECGs with mapped $\bos{s}_i$ vectors. 

During the course of the relativistic computations presented in this paper, the $D_{2\text{h}}$ point group (Table \ref{table:D2h}) was used as a finite subgroup for both $O(3)$ and $D_{\infty\text{h}}$, since the relevant ($\Sigma$ and $\Pi$) spatial parts could be well represented (in separate irreps) in this subgroup. (It was also practical that the singlet and $x,y,z$ triplet representation of the two-electron spin functions also transform according to irreps of $D_{2\text{h}}$ (see Sec.~III of the main text).) 

The effect of the $D_{2\text{h}}$ symmetry operations on the $\bos{s}_i$ shift vector of the fECG spatial functions amounts to swapping the sign and/or interchanging components.

The $D_{2\text{h}}$ irreps corresponding to the non-relativistic states computed in this work are collected in Table \ref{tab:irreps}.
The lowest-energy $S^\texte$, $P^\texto$, $\Sigma_\text{g}^+$, and $\Pi_\text{g}$ non-relativistic states were obtained as ground state in the corresponding irrep.

The $P^\texte$ and $\Sigma_\text{g}^-$ states appear as excited states in their corresponding $D_{2\text{h}}$ irrep, and the lower-energy states in their $D_{2\text{h}}$ irrep correspond to some atomic and molecular state of different symmetry in $O(3)$ and $D_{\infty\text{h}}$, respectively. 
For an efficient generation of a non-relativistic parameter set, we have considered the $D_{4\text{h}}$ point group for these states to perform the parameter optimization (Table \ref{table:D4h}), and then, `expanded' the $D_{4\text{h}}$-adapted fECG basis to a corresponding $D_{2\text{h}}$ fECG basis by explicitly generating a second series of basis functions (by doubling the fECG basis) with $\bos{s}$ vectors rotated by $\pi/4$ around the $z$ axis.    

In what follows, we collect the technical details to complete the documentation of the computations reported in the paper.

\subsection{\texorpdfstring{$O\mathrm{(3)}$}{} point group}

The $O(3)$ rotation-inversion group is the direct product of the three-dimensional rotation group $SO\mathrm{(3)}$ and the $C_\mathrm{I}$ group composed of the unity and the inversion operators.
The irreps of $O(3)$ can be classified according to the quantum number of the angular momentum ($L = 0, 1, 2, \dots$), the projection of the angular momentum on the $z$-axis ($M_L = -L, -L+1, \dots, L-1, L$), and the parity ($p=\pm 1$). 

\subsubsection{\texorpdfstring{$S^\texte$}{}}

$S^\texte$ labels the totally symmetric irrep corresponding to the quantum numbers $L=0$, $M_L=0$, and $p=+1$. An ECG centered at the origin, \emph{i.e.,} the s-vectors set to zero $\bs^{(0)}=0$, transforms according to this irrep. 
Then, of course, this function is invariant to the action of all symmetry operators in $D_{2\text{h}}$,  and projection onto the totally symmetric $A_\mathrm{g}$ irrep of $D_{2\text{h}}$ is trivially obtained as
\begin{align}
  \Theta^{A_\mathrm{g}}_{D_{2\text{h}}}( \bos{A},\bos{s}^{(0)})
  =
  P^{A_\mathrm{g}}_{D_{2\text{h}}} \Theta (\bos{A},\bos{s}^{(0)})
  =  
  \Theta(\bos{A}, \bos{s}^{(0)}) 
  =
  \exp\left[%
    -\bos{r}^\text{T} \underline{\bos{A}}\bos{r}
  \right] \; ,
\end{align}
where $P^{A_\mathrm{g}}_{D_{2\text{h}}}$ is projector to the totally symmetric irrep of $D_{2\text{h}}$. This form is also identical with Eq.~(\ref{eq:polangpref}) for $(L,M_L,p)=(0,0,+1)$ with no polynomial prefactors, and for the present atomic computations, it provided an appropriate basis representation.

\subsubsection{\texorpdfstring{$P^\texto$}{}}

$P^\texto$ labels the three-dimensional irrep of $O(3)$ corresponding to the quantum numbers $L=1$ and $p=-1$, which is often represented by basis functions corresponding to $M_L=-1, 0, 1$ quantum numbers. In the present work, we will instead instead the $x,y,z$ real representation. In the non-relativistic computations, we have parameterized $z$-type ($M_L=0$) functions, which correspond to the $z$-axis component, $P_z^\texto$, and has a cylindrical symmetry about the $z$ axis.

We described these functions by using fECGs with the $\bos{s}^{(z)}$ vectors `fixed' to the $z$ axis, \emph{i.e.,} $\bos{s}^{(z)\text{T}}=(0,0,s_1,0,0,s_2)$ with 
$s_1$ and $s_2 \in \mathbb{R}$ corresponding to particle `one' and particle `two'. 
This special parameterization of the fECG, translates to simple transformation of the $\bos{s}$ vectors by $D_{2 \mathrm{h}}$ operations, they are either invariant to the action of the symmetry operator or change sign.  
A $P_z^\texto$-type function transform according to the $B_{1\mathrm{u}}$ irrep of $D_{2 \mathrm{h}}$ (Table~\ref{tab:irreps}) leading to the following form for the  projected fECG function,
\begin{align}
  \Theta^{B_{1 \mathrm{u}}}_{D_{2\mathrm{h}}} (\bos{A},\bos{s}^{(z)} )
  =
  P^{B_{1\mathrm{u}}}_{D_{2\text{h}}} \Theta(\bos{A},\bos{s}^{(z)})
  =
  \frac{1}{\sqrt{2}}
  \left\{  
    \Theta(\bos{A},\bos{s}^{(z)})
    -
    \Theta(\bos{A},-\bos{s}^{(z)})
  \right\} \ .
\end{align}

To perform the relativistic computations, not only the $z$-, but also the $x$- and the $y$-component functions are required. The $P_x^\texto$ and $P_y^\texto$ real combinations were obtained by mapping the $\bos{s}$ vectors to the $x$ and $y$ axis, resulting in 
 $\bos{s}^{(x)\text{T}}=(s_1,0,0,s_2,0,0)$ and $\bos{s}^{(y)\text{T}}=(0,s_1,0,0,s_2,0)$, respectively.
The resulting $P_x^\texto$-adapated fECG (corresponding to the $B_{3\mathrm{u}}$ irrep of $D_{2\text{h}}$) reads as
\begin{align}
  \Theta^{B_{3\mathrm{u}}}_{D_{2\text{h}}} (\bos{A},\bos{s}^{(x)})
  =
  \frac{1}{\sqrt{2}}
  \left\{
    \Theta(\bos{A},\bos{s}^{(x)})
    -
    \Theta(\bos{A},-\bos{s}^{(x)})
  \right\} \; ,
\end{align}
and, the $P_y^\texto$ component is 
\begin{align}
   \Theta^{B_{2\mathrm{u}}}_{D_{2\text{h}}}(\bos{A},\bos{s}^{(y)})
  =
  \frac{1}{\sqrt{2}}
  \left\{ 
    \Theta(\bos{A},\bos{s}^{(y)})
    -
    \Theta(\bos{A},\bos{s}^{(y)})
  \right\}\ .
  \end{align}

\subsubsection{\texorpdfstring{$P^e$}{}  \label{sec:unnatpar}}

$P^\texte$ labels the three-dimensional irrep of $O$(3) corresponding to the quantum numbers $L=1$ and $p=+1$, which is often represented by basis functions with $M_L=-1, 0, 1$ quantum numbers. 
These type of functions exist for systems with at least two particles, since an even-parity function with $L=1$ angular momentum can be `generated' only by coupling two $L=1$, $p=-1$ functions, \emph{i.e.,} by coupling two natural-parity one-particle functions. 
The resulting angular symmetry can be described by coupling spherical harmonics functions, $Y_{L,M_L}(\hat{r}_i)$  \cite{suzukiStochasticVariationalApproach1998,komasaStatesMolecularHydrogen2008,suzukiExcitedStatesPositronium2000,bromleyPositronicComplexesUnnatural2007}, {\it e.g.}, for $M_L=0$,
\begin{align}
\eta_\mathrm{sph}^{(1,0,1)}\left( \hat{r}_1, \hat{r}_2\right)
=
\frac{1}{\sqrt{2}}
\left[%
  Y_{1,-1}(\hat{r}_1)Y_{1,1}(\hat{r}_2)
  -
  Y_{1,1}(\hat{r}_1)Y_{1,-1}(\hat{r}_2)  \right] \ .
  \label{eq:unnatspherpart}
\end{align}
For the present work, the real, $x,y,z$-representation is more useful,
\begin{align}
    Q_{1,x}(\hat{r})&= \frac{1}{\sqrt{2}} \left[ Y_{1,-1}(\hat{r}) + Y_{1,1}(\hat{r}) \right] \ , \\
    Q_{1,y}(\hat{r})&= \frac{\iim}{\sqrt{2}} \left[ Y_{1,-1}(\hat{r}) - Y_{1,1}(\hat{r}) \right] \ ,
\end{align}
and thus, $\eta^{(1,0,1)}$ can also be written as
\begin{align}
\eta_\mathrm{sph}^{(1,0,1)}\left( \hat{r}_1, \hat{r}_2\right)
=
\frac{1}{\sqrt{2}}
\left[%
  Q_{1,x}(\hat{r}_1)Q_{1,y}(\hat{r}_2)
  -
  Q_{1,y}(\hat{r}_1)Q_{1,x}(\hat{r}_2)  \right] \ .
  \label{eq:exactunnatspherpart}
\end{align}
To describe the  $Q_{1,x}(\hat{r}_1)Q_{1,y}(\hat{r}_2)$ product, we consider $\bos{s}^{(x|y)\text{T}}=(s_1,0,0,0,s_2,0)$.
Then, $P^\texte_z$-type spatial functions are generated by using fECGs with the special $\bos{s}^{(x|y)\text{T}}$ parameterization projected onto the $A_{2\text{g}}$ irrep of the $D_{4\text{h}}$ group,     
\begin{align}
 \label{eq:unnatapproxa2g}
  \Theta^{A_{2 \mathrm{g}}}_{D_{4\mathrm{h}}} (\bos{A},\bos{s}^{(x|y)})
  =
  P^{A_{2\mathrm{g}}}_{D_{4\text{h}}} &\Theta(\bos{A},\bos{s}^{(x|y)}) =
   \\ \nonumber
   \frac{1}{2\sqrt{2}}
  {\Big \{ } 
     &\Theta(\bos{A},\bos{s}^{(x|y)})
    -\Theta(\bos{A},\bos{s}^{(\bar{x}|y)})
    +\Theta(\bos{A},-\bos{s}^{(x|y)})
    -\Theta(\bos{A},-\bos{s}^{(\bar{x}|y)}) \\ \nonumber
   -&\Theta(\bos{A},\bos{s}^{(y|x)})
    +\Theta(\bos{A},\bos{s}^{(\bar{y}|x)})
    -\Theta(\bos{A},-\bos{s}^{(y|x)})
    +\Theta(\bos{A},-\bos{s}^{(\bar{y}|x)})
  {\Big \} } \ , 
\end{align}
where $\bos{s}^{(\bar{x}|y)\text{T}}=(-s_1,0,0,0,s_2,0)$, $\bos{s}^{(y|x)\text{T}}=(0,s_2,0,s_1,0,0)$ , and $\bos{s}^{(\bar{y}|x)\text{T}}=(0,-s_2,0,s_1,0,0)$.
$Q_{1,x}(\hat{r}_1)Q_{1,y}(\hat{r}_2)$ of Eq.~(\ref{eq:exactunnatspherpart}) is represented by the first four terms in Eq.~(\ref{eq:unnatapproxa2g}), and $Q_{1,y}(\hat{r}_1)Q_{1,x}(\hat{r}_2)$ of Eq.~(\ref{eq:exactunnatspherpart}) is by the the last four terms of Eq.~(\ref{eq:unnatapproxa2g}).

During the relativistic computations presented in this work, the $D_{2 \text{h}}$ point group is used.
A $\Theta^{A_{2 \mathrm{g}}}_{D_{4\mathrm{h}}} (\bos{A},\bos{s}^{(x|y)})$ function can be `expanded' to $D_{2\text{h}}$
by considering, 
\begin{align}
  \Theta^{A_{2\mathrm{g}}}_{D_{4\mathrm{h}}} (\bos{A},\bos{s}^{(x|y)} )
  &=
   P^{A_{2\mathrm{g}}}_{D_{4\text{h}}} \Theta(\bos{A},\bos{s}^{(x|y)}) 
   =
   P^{B_{1\mathrm{g}}}_{D_{2\text{h}}} 
   \left\{%
      \Theta(\bos{A},\bos{s}^{(x|y)})
     -\Theta(\bos{A},\bos{s}^{(y|x)}) 
   \right\} \ .
   \label{eq:expandanti}
\end{align}
By `expansion' of the basis, we mean that in practice, every $\Theta(\bos{A},\bos{s}^{(x|y)})$ fECG function is copied 
with a mapped s-vector parameterization to $\Theta(\bos{A},\bos{s}^{(y|x)})$. So, the basis size is doubled, and the 
Eq.~\eqref{eq:expandanti} antisymmetric combination is obtained by diagonalization of the Hamiltonian in the expanded basis set. 
     
The $P^\texte_x$- and $P^\texte_y$-type basis functions can be generated from the corresponding $P^\texte_z$-type function by mapping (`copying') the $s^{(x|y)}$ vectors onto the $yz$ and $xz$ planes, respectively, 
\begin{align}
  \Theta^{E_{\mathrm{g,x}}}_{D_{4\mathrm{h}}}(\bos{A},\bos{s}^{(y|z)})
  &=
    P^{B_{3\mathrm{g}}}_{D_{2\text{h}}} 
    \left\{%
      \Theta(\bos{A},\bos{s}^{(y|z)})-\Theta(\bos{A},\bos{s}^{(z|y)}) 
    \right\} \ , \\
  \Theta^{E_{ \mathrm{g},y}}_{D_{4\mathrm{h}}}(\bos{A},\bos{s}^{(x|z)})
  &=
    P^{B_{2\mathrm{g}}}_{D_{2\text{h}}} 
    \left\{%
      \Theta(\bos{A},\bos{s}^{(x|z)})-\Theta(\bos{A},\bos{s}^{(z|x)})
    \right\} \ .
 \end{align}

\subsection{\texorpdfstring{$D_{\infty \mathrm{h}}$}{} point group \label{sec:dinfth}}

The $D_{\infty \mathrm{h}}$ point group includes a continuous symmetry for the one-dimensional rotation about the $z$-axis. 
Perpendicular to this rotation axis, the group includes also plane symmetry.
Its irreps can be classified according to the angular momentum projection onto the $z$-axis ($M_L=\pm L$ with $L=0, 1, 2, \dots$), the parity of the inversion symmetry ($p_I=\pm1$),  and the parity of the plane symmetry parallel with the $z$-axis for $L=0$ states ($p_P=\pm 1$).

\subsubsection{\texorpdfstring{$\Sigma_\mathrm{g}^+$}{} }
$\Sigma_\mathrm{g}^+$ labels the totally symmetric irrep with the `quantum numbers' (characters) $M_L=0$, $p_I=1$, and $p_P=1$.
The ECG basis functions corresponding to this irrep can be generated by using s-vectors restricted to the $z$-axis,  \emph{i.e.,}  $\bos{s}^{(z)}=(0,0,s_1,0,0,s_2),\ s_1,s_2\in\mathbb{R}$, and then, 
with this restriction, the symmetry projection in $D_{\infty \text{h}}$ or equivalently in its subgroup (used in this work), $D_{2\mathrm{h}}$, results in the same simple expression,
\begin{align}
  \Theta^{\Sigma_{\mathrm{g}}^+}_{D_{\infty\mathrm{h}}}(\bos{A},\bos{s}^{(z)})
  =
  P^{\Sigma_{\mathrm{g}}^+}_{D_{\infty\text{h}}} \Theta(\bos{A},\bos{s}^{(z)})
  =
  \frac{1}{\sqrt{2}}
  \left\{  
    \Theta(\bos{A},\bos{s}^{(z)})
    +
    \Theta(\bos{A},-\bos{s}^{(z)})
  \right\} \ ,
\end{align}

\begin{align}
  \Theta^{A_{\mathrm{g}}}_{D_{2\mathrm{h}}} (\bos{A},\bos{s}^{(z)})
  =
  P^{A_{\mathrm{g}}}_{D_{2\text{h}}} \Theta(\bos{A},\bos{s}^{(z)})
  =
  \frac{1}{\sqrt{2}}
  \left\{  
    \Theta(\bos{A},\bos{s}^{(z)})
    +
    \Theta(\bos{A},-\bos{s}^{(z)})
  \right\} \ . 
\end{align}

\subsubsection{\texorpdfstring{$\Pi_\text{g}$}{} }
$\Pi_\text{g}$ labels a two-dimensional irrep with $M_L=\pm1$, and $p_I=1$. For computational convenience, 
instead of using $M_L=1$ and $M_L=-1$-type basis functions, we use the $x,y$ representation, \emph{i.e.,} $\Pi_{\text{g},x}=(\Pi_{\text{g},-1}+\Pi_{\text{g},1})/\sqrt{2}$ and  $\Pi_{\text{g},y}=\iim (\Pi_\mathrm{g,-1}-\Pi_\mathrm{g,-1})/\sqrt{2}$ states. 
Since the $\Pi_{\text{g},y}$ state has a nodal surface in the $xz$-plane, we can describe the $\Pi_{\text{g},x}$ state separately by restricting the ECGs shift vectors (`centers') to this plane,   \emph{i.e.,} by the special parameterization, $\bos{s}^{(xz)\text{T}}=(s_1,0,s_2,s_3,0,s_4),\  s_1,s_2,s_3,s_4\in\mathbb{R}$. 
We use this parameterization, and carry out subgroup projection onto the corresponding 
$B_{2\text{g}}$ irrep of the $D_{2\text{h}}$ group (Table~\ref{tab:irreps}), 
\begin{align}
  \Theta^{B_{3\mathrm{g}}}_{D_{2\mathrm{h}}} (\bos{A},\bos{s}^{(xz)})
  =
  &P^{B_{3\mathrm{g}}}_{D_{2\text{h}}} \Theta(\bos{A},\bos{s}^{(xz)})
  = \\ \nonumber 
 &  \frac{1}{2}
  \left\{  
     \Theta(\bos{A},\bos{s}^{(xz)})
    +\Theta(\bos{A},-\bos{s}^{(xz)})
    -\Theta(\bos{A},\bos{s}^{(\bar{x}z)})
    -\Theta(\bos{A},-\bos{s}^{(\bar{x}z)})
  \right\} \ ,
\end{align}
where $\bos{s}^{(\bar{x}z)\text{T}}=(-s_1,0,s_2,-s_3,0,s_4)$.

Then, the corresponding $\Pi_{\text{g},y}$-type functions (necessary for the relativistic computations) can be generated by mapping the $s^{(xz)}$ parameterization (generated and optimized for the $x$-component) onto the $yz$ plane and by carrying out subgroup projection to the corresponding $B_{2 \text{g}}$ irrep of the $D_{2\text{h}}$ group (Table~\ref{tab:irreps}), 
\begin{align}
  \Theta^{B_{2\text{g}}}_{D_{2\mathrm{h}}}(\bos{A},\bos{s}^{(yz)})
  =
  &P^{B_{2\text{g}}}_{D_{2\text{h}}} \Theta(\bos{A},\bos{s}^{(yz)})
  = \\ \nonumber 
 &  \frac{1}{2}
  \left\{  
     \Theta(\bos{A},\bos{s}^{(yz)})
    +\Theta(\bos{A},-\bos{s}^{(yz)})
    -\Theta(\bos{A},\bos{s}^{(\bar{y}z)})
    -\Theta(\bos{A},-\bos{s}^{(\bar{y}z)})
  \right\} \ . 
\end{align}

\subsubsection{\texorpdfstring{$\Sigma_\text{g}^-$}{} }

$\Sigma_\text{g}^-$ is a one-dimensional irrep with `quantum numbers' (characters) $M_L=0$, $p_I=1$, and $p_P=-1$. 
It can be considered as a `diatomic' counterpart of the $P^\texte$-type atomic functions (Sec.~\ref{sec:unnatpar}). 

We can construct the angular part as (to have obtain $p_I=1$ and $p_P=-1$) similarly to Eq.~(\ref{eq:unnatspherpart}),
\begin{align}
\eta_\mathrm{cyl}^{(0,0,1,-1)}\left( \varphi_1, \varphi_2\right)
=
\frac{1}{\sqrt{2}}
\left[%
  y_{1,-1}(\varphi_1)y_{1,1}(\varphi_2)
  -
  y_{1,1}(\varphi_1)y_{1,-1}(\varphi_2)  \right] \ 
  \label{eq:2dunnatspherpart}
\end{align}
with $y_{L,M_L}(\varphi) = e^{\iim M_L \varphi}$, $\varphi \in [0,2\pi)$. 
Instead of the $\pm 1$-representation, we use the $x,y$-representation,
\begin{align}
  q_{1,x}(\varphi)
  &= 
  \frac{1}{\sqrt{2}} \left[ y_{1,-1}(\varphi) + y_{1,1}(\varphi) \right] \ , \\
  q_{1,y}(\varphi)
  &= 
  \frac{\iim}{\sqrt{2}} \left[ y_{1,-1}(\varphi) - y_{1,1}(\varphi) \right] \ ,
\end{align}
and thus, Eq.~(\ref{eq:2dunnatspherpart}) can be written as
\begin{align}
  \eta_\mathrm{cyl}^{(0,0,1,-1)}\left( \varphi_1, \varphi_2\right)
  =
  \frac{1}{\sqrt{2}}
  \left[%
    q_{1,x}(\varphi_1)q_{1,y}(\varphi_2)
    -
    q_{1,y}(\varphi_1)q_{1,x}(\varphi_2)  
  \right] \ .
  \label{eq:2dunnatspherpartq}
\end{align}

To construct an fECG representation for the $q_{1,x}(\varphi_1)q_{1,y}(\varphi_2)$ product, 
we start by considering a special $\bos{s}^{(x|y,z)\text{T}}=(s_1,0,s_2,0,s_3,s_4),\ s_1,s_2,s_3,s_4\in\mathbb{R}$
s-vector parameterization.
Then, we carry out subgroup projection onto the corresponding  $A_{2\mathrm{g}}$ irrep in the $D_{4\text{h}}$ group (Table~\ref{tab:irreps}) for this special parameterization as
\begin{align}
 \label{eq:unnatapprox}
  \Theta^{A_{2 \text{g}}}_{D_{4\text{h}}}&(\bos{A},\bos{s}^{(x|y,z)})
  =
  P^{A_{2\text{g}}}_{D_{4\text{h}}}\Theta(\bos{A},\bos{s}^{(x|y,z)})
  = \\ \nonumber
   \hspace{0.5cm} \frac{1}{4}
  {\Big \{ } 
     &\Theta(\bos{A},\bos{s}^{(x|y,z)})
    -\Theta(\bos{A},\bos{s}^{(\bar{x}|y,z)})
    +\Theta(\bos{A},-\bos{s}^{(x|y,z)})
    -\Theta(\bos{A},-\bos{s}^{(\bar{x}|y,z)}) \\ \nonumber
    -&\Theta(\bos{A},\bos{s}^{(y|x,z)})
    +\Theta(\bos{A},\bos{s}^{(\bar{y}|x,z)})
    -\Theta(\bos{A},-\bos{s}^{(y|x,z)})
    +\Theta(\bos{A},-\bos{s}^{(\bar{y}|x,z)}) \\  \nonumber
    +&\Theta(\bos{A},\bos{s}^{(x|y,\bar{z})})
    -\Theta(\bos{A},\bos{s}^{(\bar{x}|y,\bar{z})})
    +\Theta(\bos{A},-\bos{s}^{(x|y,\bar{z})})
    -\Theta(\bos{A},-\bos{s}^{(\bar{x}|y,\bar{z})}) \\ \nonumber
    -&\Theta(\bos{A},\bos{s}^{(y|x,\bar{z})})
    +\Theta(\bos{A},\bos{s}^{(\bar{y}|x,\bar{z})})
    -\Theta(\bos{A},-\bos{s}^{(y|x,\bar{z})})
    +\Theta(\bos{A},-\bos{s}^{(\bar{y}|x,\bar{z})})
  {\Big \} } 
\end{align}
where 
\begin{align}
  \bos{s}^{(\bar{x}|y,z)\text{T}}& =(-s_1,0,s_2,0,s_3,s_4) \ , \\ \bos{s}^{(y|x,z)\text{T}}&=(0,s_3,s_2,s_1,0,s_4) \ , \\ \bos{s}^{(\bar{y}|x,z)\text{T}}&=(0,-s_3,s_2,s_1,0,s_4) \ , \\ 
  \bos{s}^{(x|y,\bar{z})\text{T}}&=(s_1,0,-s_2,0,s_3,-s_4) \ , \\
\bos{s}^{(\bar{x}|y,\bar{z})\text{T}}&=(-s_1,0,-s_2,0,s_3,-s_4) \ , \\ \bos{s}^{(y|x,\bar{z})\text{T}}&=(0,s_3,-s_2,s_1,0,-s_4) \ , \\ \bos{s}^{(\bar{y}|x,\bar{z})\text{T}}&=(0,-s_3,-s_2,s_1,0,-s_4) \ . 
\end{align}
This representation is used to generate and optimize an fECG basis set by minimization of the non-relativistic energy. For computations carried out in this work (H$_2$), the relevant $\Sigma_\text{g}^-$ state is obtained as the ground state in the $A_{2\text{g}}$ irrep of $D_{4\text{h}}$.

During the course of the relativistic computations, we use $D_{2 \mathrm{h}}$ subgroup projection, and thus, we `expand' the $A_{2\text{g}}$ ($D_{4\text{h}}$) symmetry-adapted fECG basis set to the corresponding $B_{1\mathrm{g}}$ ($D_{2\text{h}}$) according to the relation
\begin{align}
  \Theta^{A_{2\mathrm{g}}}_{D_{4\mathrm{h}}}(\bos{A},\bos{s}^{(x|y,z)})
  &=
   P^{A_{2\mathrm{g}}}_{D_{4\text{h}}} \Theta(\bos{A},\bos{s}^{(x|y,z)}) =
    P^{B_{1\mathrm{g}}}_{D_{2\text{h}}} %
    \left\{%
       \Theta(\bos{A},\bos{s}^{(x|y,z)})
      -\Theta(\bos{A},\bos{s}^{(y|x,z)})
    \right\} \ .
    \label{eq:expandanti2}
\end{align}
So, in practice, we copy every $\Theta\left[\bos{A},\bos{s}^{(x|y,z)}\right]$ basis function, while mapping the s-parameterization to $\Theta\left[\bos{A},\bos{s}^{(y|x,z)}\right]$. This doubles the basis size, and the anti-symmetric combination in Eq.~\eqref{eq:expandanti2} is obtained by diagonalization.

\begin{table}[ht]
    \centering
    \caption{%
      Collection of irreducible representations used during this work. 
    }
    \label{tab:irreps}
    \begin{tabular}{@{}ccllll@{}}
       \hline \hline
              spin  & linear functions &  &  & &   \\
        functions & and rotations & $O$(3) & $D_{\infty \mathrm{h}}$ & $D_{4\mathrm{h}}$&  $D_{2\mathrm{h}}$   \\ \hline
        $S$   & --    & $S^\texte$   & $\Sigma_\mathrm{g}^+$ & $A_{1\mathrm{g}}$    & $A_\mathrm{g}$    \\ 
        $T_z$ & $R_z$ & $P_z^\texte$ & $\Sigma_\mathrm{g}^-$ & $A_{2\mathrm{g}}$    & $B_{1\mathrm{g}}$ \\
        $T_y$ & $R_y$ & $P_y^\texte$ & $\Pi_{\mathrm{g},y}$  & $E_{\mathrm{g},y}$   & $B_{2\mathrm{g}}$ \\
        $T_x$ & $R_x$ & $P_x^\texte$ & $\Pi_{\mathrm{g},x}$  & $E_{\mathrm{g},x}$   & $B_{3\mathrm{g}}$ \\
        --    & $z$   & $P_z^\texto$ & $\Sigma_\mathrm{u}^+$ & $A_{2\mathrm{u}}$    & $B_{1\mathrm{u}}$ \\
        --    & $y$   & $P_y^\texto$ & $\Pi_{\mathrm{u},y}$  & $E_{{\mathrm{u},y}}$ & $B_{2\mathrm{u}}$ \\
        --    & $x$   & $P_x^\texto$ & $\Pi_{\mathrm{u},x}$  & $E_{{\mathrm{u},x}}$ & $B_{3\mathrm{u}}$ \\ 
        \hline \hline
    \end{tabular}
\end{table}

 \begin{table}[ht]
     \centering
          \caption{Character table of the $D_{2\text{h}}$ point group.}
     \label{table:D4h}
     \begin{tabular}{@{}c|cccccccc|cc@{}}
       \hline \hline
        & &  &  &  &  & &  &  & spin & linear functions\\ \
          & $E$ & $C_{2z}$ &  $C_{2y}$ &  $C_{2x}$ & $i$ & $\sigma_{xy}$ &  $\sigma_{xz}$ &  $\sigma_{yz}$ & functions & and rotations\\ \hline
        $A_\mathrm{g}$ & \hspace{0.3cm}1 & \hspace{0.3cm}1 & \hspace{0.3cm}1 & \hspace{0.3cm}1 & \hspace{0.3cm}1 & \hspace{0.3cm}1 & \hspace{0.3cm}1 & \hspace{0.3cm}1 & $S$ & \\
        $B_\mathrm{1g}$ & \hspace{0.3cm}1 & \hspace{0.3cm}1 & $-$1 & $-$1 & \hspace{0.3cm}1 & \hspace{0.3cm}1 & $-$1 & $-$1 & $T_z$ &  $R_z$ \\
        $B_\mathrm{2g}$ & \hspace{0.3cm}1 & $-$1 & \hspace{0.3cm}1 & $-$1 & \hspace{0.3cm}1 & $-$1 & \hspace{0.3cm}1 & $-$1 & $T_y$ & $R_y$ \\
        $B_\mathrm{3g}$ & \hspace{0.3cm}1 & $-$1 & $-$1 & \hspace{0.3cm}1 & \hspace{0.3cm}1 & $-$1 & $-$1 & \hspace{0.3cm}1 & $T_x$ & $R_x$ \\
        $A_\mathrm{u}$  & \hspace{0.3cm}1 &  \hspace{0.3cm}1 &  \hspace{0.3cm}1 & \hspace{0.3cm}1 & $-$1& $-$1 & $-$1 & $-$1&  &  \\
        $B_\mathrm{1u}$ & \hspace{0.3cm}1 &  \hspace{0.3cm}1 & $-$1 & $-$1& $-$1& $-$1 &  \hspace{0.3cm}1 & \hspace{0.3cm}1 &  & $z$ \\
        $B_\mathrm{2u}$ & \hspace{0.3cm}1 & $-$1 &  \hspace{0.3cm}1 & $-$1& $-$1&  \hspace{0.3cm}1 & $-$1 & \hspace{0.3cm}1 &  & $y$ \\
        $B_\mathrm{3u}$ & \hspace{0.3cm}1 & $-$1 & $-$1 &  \hspace{0.3cm}1& $-$1&  \hspace{0.3cm}1 &  \hspace{0.3cm}1 & $-$1&  & $x$ \\
       \hline \hline        
     \end{tabular}
 \end{table}
 
  \begin{table}[ht]
     \centering
          \caption{Character table of the $D_{4\text{h}}$ point group.}
     \label{table:D2h}
     \begin{tabular}{@{}c|cccccccccc|cc@{}}
       \hline \hline
       & &  &  &  & & &  &  &   &  & spin & linear functions \\ 
          & $E$ & $2C_{4z}$ &  $C_{2}$ &  $2C_{2}'$ & $2C_{2}''$ & $i$ & $2S_{4}$ & $\sigma_\mathrm{h}$ &  $2\sigma_\mathrm{v}$ &  $2\sigma_\mathrm{d}$ & functions & and rotations \\ \hline
        $A_\mathrm{1g}$ & \hspace{0.3cm}1 & \hspace{0.3cm}1 & \hspace{0.3cm}1 & \hspace{0.3cm}1 & \hspace{0.3cm}1 & \hspace{0.3cm}1 & \hspace{0.3cm}1 & \hspace{0.3cm}1 & \hspace{0.3cm}1 & \hspace{0.3cm}1 & $S$ &  \\
        $A_\mathrm{2g}$ & \hspace{0.3cm}1 & \hspace{0.3cm}1 & \hspace{0.3cm}1 & $-$1 & $-$1 & \hspace{0.3cm}1 & \hspace{0.3cm}1 & \hspace{0.3cm}1 & $-$1 & $-$1 & $T_z$ & $R_z$ \\
        $B_\mathrm{1g}$ & \hspace{0.3cm}1 & $-$1 & \hspace{0.3cm}1 & \hspace{0.3cm}1 & $-$1 & \hspace{0.3cm}1 & $-$1 & \hspace{0.3cm}1 & \hspace{0.3cm}1 & $-$1 &  &    \\
        $B_\mathrm{2g}$ & \hspace{0.3cm}1 & $-$1 & \hspace{0.3cm}1 & $-$1 & \hspace{0.3cm}1 & \hspace{0.3cm}1 & $-$1 & \hspace{0.3cm}1 & $-$1 & \hspace{0.3cm}1&  &    \\
        $E_\mathrm{g}$  & \hspace{0.3cm}2 & \hspace{0.3cm}0 &  $-$2 & \hspace{0.3cm}0 & \hspace{0.3cm}0 & \hspace{0.3cm}2 & \hspace{0.3cm}0 & $-$2 & \hspace{0.3cm}0 & \hspace{0.3cm}0 & $T_x$, $T_y$ & $R_x$, $R_y$   \\
        $A_\mathrm{1u}$ & \hspace{0.3cm}1 &  \hspace{0.3cm}1 &  \hspace{0.3cm}1 & \hspace{0.3cm}1 & \hspace{0.3cm}1 & $-$1 & $-$1 & $-$1& $-$1& $-$1&  &   \\
        $A_\mathrm{2u}$ & \hspace{0.3cm}1 &  \hspace{0.3cm}1 &  \hspace{0.3cm}1 & $-$1 & $-$1& $-$1 & $-$1 & $-$1& \hspace{0.3cm}1& \hspace{0.3cm}1&  &   \\
        $B_\mathrm{1u}$ & \hspace{0.3cm}1 &  $-$1 &  \hspace{0.3cm}1 &  \hspace{0.3cm}1 & $-$1& $-$1 &  \hspace{0.3cm}1 & $-$1 & $-$1& \hspace{0.3cm}1& & $z$  \\
        $B_\mathrm{2u}$ & \hspace{0.3cm}1 &  $-$1 &  \hspace{0.3cm}1 & $-$1&   \hspace{0.3cm}1& $-$1 &  \hspace{0.3cm}1 & $-$1 &  \hspace{0.3cm}1& $-$1& &   \\
        $E_\mathrm{u}$  & \hspace{0.3cm}2 & \hspace{0.3cm}0 &  $-$2 & \hspace{0.3cm}0 & \hspace{0.3cm}0 & $-$2 & \hspace{0.3cm}0 & \hspace{0.3cm}2 & \hspace{0.3cm}0 & \hspace{0.3cm}0 &  & $x$, $y$   \\
       \hline \hline        
     \end{tabular}
 \end{table}

\clearpage
\section{Convergence Tables}

\begin{table}[ht]
    \centering
    \caption{Convergence of the non-relativistic energy, in n$\Eh$, for the lowest $^3P^\texte$ (unnatural-parity) state of helium isoelectronic systems. The calculation is performed with quadruple precision arithmetic.}
    \label{tab:unnatparityatom}
    \begin{tabular}{@{}lccc@{}}
       \hline \hline \\[-0.35cm]
        \hspace{0.15cm}$N_\text{t}$ & He & Li$^+$&  Be$^{2+}$  \\ \hline
         \hspace{0.15cm}10 & $-$0.710 214 115 & $-$1.794 878 123 & $-$3.381 768 665 \\ 
         \hspace{0.15cm}25 & $-$0.710 482 214 & $-$1.796 573 761 & $-$3.382 678 261 \\
         \hspace{0.15cm}35 & $-$0.710 496 353 & $-$1.796 642 156 & $-$3.382 686 239 \\
         \hspace{0.15cm}50 & $-$0.710 499 120 & $-$1.796 644 518 & $-$3.382 706 952\\
         \hspace{0.15cm}75 & $-$0.710 499 767 & $-$1.796 647 688 &  $-$3.382 710 997 \\
         100 &$-$0.710 500 025 & $-$1.796 647 978 & $-$3.382 711 411\\
         125 &$-$0.710 500 037 & $-$1.796 648 056 & $-$3.382 711 809\\ 
         150 &  \multicolumn{1}{c}{--}     & $-$1.796 648 085 &  \multicolumn{1}{c}{--}     \\ 
         \hline
        $E_\mathrm{ref}$ \cite{hilgerAccurateNonrelativisticEnergies1996,jiangWavelengths2pnp1Pe2012} & 
        $-$0.710 500 156 & $-$1.796 648 100 & $-$3.382 712 421 \\ \hline \hline
    \end{tabular}
\end{table}

\begin{table}[ht]
  \caption{%
    The convergence of the non-relativistic energy, in $E_\mathrm{h}$, for the lowest $^3\Sigma_\mathrm{g}^-$ state of the H$_2$ molecule. 
    The label * indicates that the optimization  had to be performed with quadruple precision arithmetic. After the optimization, all energies were recomputed with quadruple precision arithmetic.}
    \label{tab:H2sigmagminus}
    \begin{tabular}{@{}l r c r c  r@{}}
       \hline \hline \\[-0.35cm]
        \multicolumn{1}{l}{\hspace{0.15cm}$N_t$} & 
        $R=1.0\ \mathrm{bohr}$ & & 
        $R=1.4\ \mathrm{bohr}$ & & 
        $R=1.5\ \mathrm{bohr}$ \\ \hline
        \hspace{0.15cm}10 & 0.324 418 916 & \hspace{0.5cm} & 0.062 185 280 & \hspace{0.5cm} &0.021 254 380 \\
        \hspace{0.15cm}20 & 0.323 812 165 & & 0.062 061 214 & & 0.020 599 255 \\
        \hspace{0.15cm}30 & 0.323 799 667 & & 0.061 978 210 & & 0.020 585 031 \\
        \hspace{0.15cm}40 & 0.323 788 467 & & 0.061 975 831 & & 0.020 572 422 \\
        \hspace{0.15cm}50 & 0.323 787 214 & & 0.061 973 043 & & 0.020 570 986 \\
        \hspace{0.15cm}70 & 0.323 785 792 & & 0.061 971 278 & & 0.020 568 854 \\
        100& 0.323 785 246 & & 0.061 970 325 & & 0.020 568 258 \\ 
        150*& \multicolumn{1}{c}{--}  & & 0.061 970 196 & & 0.020 567 962 \\
        200*& \multicolumn{1}{c}{--} & & 0.061 970 034 & & 0.020 567 916 \\ 
        \hline
        Ref. \cite{komasaStatesMolecularHydrogen2008} & 0.323 785 137  & &
\multicolumn{1}{c}{--} & & 0.020 567 890   \\
     \hline \hline
     \end{tabular}
\end{table}

\clearpage

\begin{table}[ht]
    \caption{Convergence of the non-relativistic energy, in $E_\mathrm{h}$, for the lowest $^3\Pi_\mathrm{g}$ state of the H$_2$ molecule for $R=1.4$~bohr interprotonic distance.}
    \label{tab:H2pig}
    \begin{tabular}{@{}l c r@{}}
        \hline \hline \\[-0.35cm]
        \hspace{0.15cm}$N_t$ & \hspace{0.5cm} & \multicolumn{1}{c}{$E$}  \\ \hline
        100& & $-$0.626 227 619 3 \\ 
        150& & $-$0.626 230 281 7 \\
        200& & $-$0.626 233 311 3 \\ 
        300& & $-$0.626 233 638 7 \\ 
        400& & $-$0.626 233 724 6\\ 
        500& & $-$0.626 233 888 8\\
        600& & $-$0.626 233 911 3\\\hline
        Ref. \cite{silkowskiBornOppenheimerPotentials2022} & & $-$0.626 234 005 7\\
        \hline \hline
    \end{tabular}
\end{table}

\begin{table}[ht]
    \caption{%
      Convergence of the relativistic energy, in $E_\text{h}$, for the lowest $^3P_0^\texto$ state of the helium atom (the non-relativistic energy is also shown).
      \label{tab:HetripletJ0}}
    \begin{tabular}{@{}l r c r c r c r c@{}}
         \hline \hline \\[-0.35cm]
         $N$ & \multicolumn{1}{c}{$E_\mathrm{nonrel}$} & & \multicolumn{1}{c}{$E_{\DC}^\pp$} & & \multicolumn{1}{c}{$E_{\DC \langle \text{B}\rangle}^\pp$} & &  \multicolumn{1}{c}{$E_{\DCB}^\pp$} & \\ \hline
         300 & \num{-2.13316416815973575182852} & & \num{-2.13327267470991}  & & \num{-2.13326475321} & & \num{-2.13326475469} &  \\
         400 & \num{-2.13316418322933} & & \num{-2.133272689974607} & & \num{-2.13326476851} & & \num{-2.13326476999} & \\
         500 & \num{-2.13316418887209} & & \num{-2.13327269601218} & & \num{-2.13326477456} & & \num{-2.13326477605} & \\
         600 & \num{-2.13316418934328} & & \num{-2.13327269649174}  & & \num{-2.13326477504} & & \num{-2.13326477653} & \\ 
         700 & \num{-2.1331641894474} & & \num{-2.13327269659369873531} &  & \num{-2.13326477514} & & \num{-2.13326477663} & \\ \hline
         $E-E_\mathrm{ref}$ & \num{0.0000000014} &$^\text{a}$& \multicolumn{1}{c}{--} &  & \multicolumn{1}{c}{--} & & \multicolumn{1}{c}{--} &  \\
         $E-E^{[2]}$ & \multicolumn{1}{c}{--} & & \num{0.0000000094}  &$^\text{b}\hspace{0.2cm}$ & \num{0.0000000095} &$^\text{c}\hspace{0.2cm}$ & \num{0.0000000110} & $^\text{c}\hspace{0.2cm}$  \\
         \hline \hline
    \end{tabular}
    \begin{flushleft}
    {\footnotesize
    $^\text{a}$ $E_\mathrm{nonrel}=\num{-2.1331641908}$ $\Eh$  \cite{drakeProgressHeliumFinestructure2002}. \\
    $^\text{b}$ $E_{\DC}^{[2]}=-$2.133 272 687 2~$E_\mathrm{h}$ \cite{schwartzFineStructureHelium1964a,drakeProgressHeliumFinestructure2002} is the perturbative DC-relativistic energy up to $\alpha^2\Eh$ order. \\
    $^\text{c}$ $E_{\mathrm{DC(B)}}^{[2]}=-$2.133 264 765 6~$E_\mathrm{h}$ \cite{schwartzFineStructureHelium1964a,drakeProgressHeliumFinestructure2002} is the perturbative DCB-relativistic energy up to $\alpha^2\Eh$ order. \\
    }
    \end{flushleft}
\end{table}

\begin{table}[ht]
    \caption{%
      Convergence of the relativistic energy, in $E_\text{h}$, for the lowest $^3P_1^\texto$ state of the helium atom (the non-relativistic energy is also shown).
    \label{tab:HetripletJ1}}
    \begin{tabular}{@{}l r c r c r c r c@{}}
         \hline \hline \\[-0.35cm]
         $N$ & \multicolumn{1}{c}{$E_\mathrm{nonrel}$} & & \multicolumn{1}{c}{$E_{\DC}^\pp$} & & \multicolumn{1}{c}{$E_{\DC \langle \text{B}\rangle}^\pp$} & &  \multicolumn{1}{c}{$E_{\DCB}^\pp$} & \\ \hline
         300 & \num{-2.13316416815973575182852} & & \num{-2.13327174290983}  & & \num{-2.1332692465} & & \num{-2.13326924673} &  \\
         400 & \num{-2.13316418322933} & & \num{-2.13327175816417} & & \num{-2.1332692618} & & \num{-2.133269262} & \\
         500 & \num{-2.13316418887209} & & \num{-2.13327175816417} & & \num{-2.1332692678} & & \num{-2.13326926803} & \\
         600 & \num{-2.13316418934328} & & \num{-2.13327176467952}  & & \num{-2.1332692683} & & \num{-2.13326926851} & \\ 
         700 & \num{-2.1331641894474} & & \num{-2.13327176478163} &  & \num{-2.1332692684} & & \num{-2.13326926861} & \\ \hline
         $E-E_\mathrm{ref}$ & \num{0.0000000014} &$^\text{a}$& \multicolumn{1}{c}{--} &  & \multicolumn{1}{c}{--} & & \multicolumn{1}{c}{--} &  \\
         $E-E^{[2]}$ & \multicolumn{1}{c}{--} & & \num{-0.0000000095}  &$^\text{b}\hspace{0.2cm}$ &  \num{-0.0000000097} &$^\text{c}\hspace{0.2cm}$ &  \num{-0.0000000097} & $^\text{c}\hspace{0.2cm}$  \\
         \hline \hline
    \end{tabular}
    \begin{flushleft}
    {\footnotesize
    $^\text{a}$ $E_\mathrm{nonrel}=\num{-2.1331641908}$~ $\Eh$  \cite{drakeProgressHeliumFinestructure2002}. \\
    $^\text{b}$ $E_{\DC}^{[2]}=-$2.133 271 755 3~$E_\mathrm{h}$ \cite{schwartzFineStructureHelium1964a,drakeProgressHeliumFinestructure2002} (see also caption to Table~\ref{tab:HetripletJ0}).\\
    $^\text{c}$ $E_{\mathrm{DC(B)}}^{[2]}=-$2.133 269 258 9~$E_\mathrm{h}$ \cite{schwartzFineStructureHelium1964a,drakeProgressHeliumFinestructure2002} 
    (see also caption to Table~\ref{tab:HetripletJ0}).
    }
    \end{flushleft}
\end{table}

\begin{table}[ht]
    \caption{%
      Convergence of the relativistic energy, in $E_\text{h}$, for the lowest $^3P_2^\texto$ state of the helium atom (the non-relativistic energy is also shown).
      \label{tab:HetripletJ2}}
    \begin{tabular}{@{}l r c r c r c r c@{}}
         \hline \hline \\[-0.35cm]
         $N$ & \multicolumn{1}{c}{$E_\mathrm{nonrel}$} & & \multicolumn{1}{c}{$E_{\DC}^\pp$} & & \multicolumn{1}{c}{$E_{\DC \langle \text{B}\rangle}^\pp$} & &  \multicolumn{1}{c}{$E_{\DCB}^\pp$} & \\ \hline
         300 & \num{-2.13316416815973575182852} & & \num{-2.13326987952853}  & & \num{-2.1332695992} & & \num{-2.1332695992} &  \\
         400 & \num{-2.13316418322933} & & \num{-2.13326989476298} & & \num{-2.1332696144}  & &  \num{-2.1332696144} & \\
         500 & \num{-2.13316418887209} & & \num{-2.13326990079622} & & \num{-2.1332696204} & & \num{-2.1332696205} & \\
         600 & \num{-2.13316418934328} & & \num{-2.13326990127569} & & \num{-2.1332696209} & &  \num{-2.1332696209} & \\ 
         700 & \num{-2.1331641894474} & & \num{-2.13326990137815}  &  &  \num{-2.133269621} &  &  \num{-2.133269621} & \\ \hline
         $E-E_\mathrm{ref}$ & \num{0.0000000014} &$^\text{a}$& \multicolumn{1}{c}{--} &  & \multicolumn{1}{c}{--} & & \multicolumn{1}{c}{--} &  \\
         $E-E^{[2]}$ & \multicolumn{1}{c}{--} & & \num{0.0000000099}  &$^\text{b}\hspace{0.2cm}$ & \num{0.0000000099} &$^\text{c}\hspace{0.2cm}$ & \num{0.0000000099} & $^\text{c}\hspace{0.2cm}$  \\
         \hline \hline
    \end{tabular}
    \begin{flushleft}
    {\footnotesize
    $^\text{a}$ $E_\mathrm{nonrel}=\num{-2.1331641908}$~$\Eh$  \cite{drakeProgressHeliumFinestructure2002}. \\
    $^\text{b}$ $E_{\DC}^{[2]}=-$2.133 269 891 5~$E_\mathrm{h}$ \cite{schwartzFineStructureHelium1964a,drakeProgressHeliumFinestructure2002}
    (see also caption to Table~\ref{tab:HetripletJ0}).
    \\
    $^\text{c}$ $E_{\mathrm{DC(B)}}^{[2]}=-$2.133 269 611 1~$E_\mathrm{h}$ \cite{schwartzFineStructureHelium1964a,drakeProgressHeliumFinestructure2002} 
    (see also caption to Table~\ref{tab:HetripletJ0}).
    }
    \end{flushleft}
\end{table}

\begin{table}[ht]
    \caption{%
      Convergence of the relativistic energy for the 1 $^1S^\texte_0$ (ground) state of the helium atom considering only singlet basis states in the basis set expansion (the non-relativistic energy is also shown).
    \label{tab:Hesinglet}}
    \begin{tabular}{@{}l r c r c r c r c r@{}}
         \hline \hline \\[-0.35cm]
         $N$ & \multicolumn{1}{c}{$E_\mathrm{nonrel}$} & & \multicolumn{1}{c}{$E_{\DC}^\pp$}  &&  \multicolumn{1}{c}{$E_{\DC \langle \text{B}\rangle}^\pp$} & &  \multicolumn{1}{c}{$E_{\DC \mathcal{B}_2}^\pp$} &&  \multicolumn{1}{c}{$E_{\DCB}^\pp$} \\ \hline
         400 & \num{-2.9037243763469788859765685} & & \num{-2.90385663130197} & &\num{-2.90382803153868} & & \num{-2.9038281212}  & &  \num{-2.90382812070168} \\
         500 & \num{-2.9037243765856652899515211174730211} & & \num{-2.90385663154708} & & \num{-2.9038280319} & & \num{-2.9038281215} & & \num{-2.9038281215} \\
         700 & \num{-2.9037243766101616948560604214435443} & & \num{-2.90385663156026} & & \num{-2.9038280319} & & \num{-2.9038281215} & &  \num{-2.9038281215} \\ \hline
         Ref. \cite{drakeHighPrecisionCalculations2006} & \num{-2.9037243770} & & \multicolumn{1}{c}{--} &  & \multicolumn{1}{c}{--} & & \multicolumn{1}{c}{--} & & \multicolumn{1}{c}{--} \\
         \hline \hline
    \end{tabular}
\end{table}

\begin{table}[ht]
        \caption{The convergence of the energy for the 2 $^1S^\texte_0$ state of the helium atom considering only singlet basis states in the basis set expansion (the non-relativistic energy is also shown).
    \label{tab:He2Ssinglet}}
    \begin{tabular}{@{}l r c r c r c r c r@{}}
         \hline \hline \\[-0.35cm]
         $N$ & \multicolumn{1}{c}{$E_\mathrm{nonrel}$} & & \multicolumn{1}{c}{$E_{\DC}^\pp$}  &&  \multicolumn{1}{c}{$E_{\DC \langle \text{B}\rangle}^\pp$} & &  \multicolumn{1}{c}{$E_{\DC \mathcal{B}_2}^\pp$} &&  \multicolumn{1}{c}{$E_{\DCB}^\pp$} \\ \hline
         400 & \num{-2.1459740441729127091718964948086068} & & \num{-2.14608478961019} & &\num{-2.14608235381559041547} & & \num{-2.14608236120223807362633}  & &  \num{-2.14608236116547} \\
         500 & \num{-2.1459740456542046871390994056127965} & & \num{-2.1460847910886} & & \num{-2.14608235552888210375} & & \num{-2.14608236295824556114571} & & \num{-2.14608236292084} \\
         600 & \num{-2.1459740457171649907763821829576045} & & \num{-2.14608479115703} & & \num{-2.14608235559875726843} & & \num{-2.14608236302844022664969} & &  \num{-2.14608236299065} \\ \hline
         Ref. \cite{drakeHighPrecisionCalculations2006} & \num{-2.14597404605} & & \multicolumn{1}{c}{--} &  & \multicolumn{1}{c}{--} & & \multicolumn{1}{c}{--} & & \multicolumn{1}{c}{--} \\
         \hline \hline
    \end{tabular}
\end{table}

\begin{table}[ht]
        \caption{The convergence of the energy for the 1 $^1S^\texte_0$ (ground) state of the Li$^+$ ion considering only singlet basis states in the basis set expansion (the non-relativistic energy is also shown).
    \label{tab:Lipsinglet}}
    \begin{tabular}{@{}l r c r c r c r c r@{}}
         \hline \hline \\[-0.35cm]
         $N$ & \multicolumn{1}{c}{$E_\mathrm{nonrel}$} & & \multicolumn{1}{c}{$E_{\DC}^\pp$}  &&  \multicolumn{1}{c}{$E_{\DC \langle \text{B}\rangle}^\pp$} & &  \multicolumn{1}{c}{$E_{\DC \mathcal{B}_2}^\pp$} &&  \multicolumn{1}{c}{$E_{\DCB}^\pp$} \\ \hline
         400 & \num{-7.2799134134} & & \num{-7.2806988997519} & & \num{-7.2805409774}  & & \num{-7.2805414447} & &  \num{-7.28054144233511} \\
         500 & \num{-7.2799134108594669712033464747946709} & & \num{-7.28069890004356} & & \num{-7.2805409782} & & \num{-7.2805414457} & & \num{-7.28054144333511} \\
         600 & \num{-7.2799134117308383906674862373620272} & & \num{-7.28069890176949} & & \num{-7.2805409800} & & \num{-7.2805414474} & &  \num{-7.28054144500112} \\ \hline
         Ref. \cite{drakeHighPrecisionCalculations2006} & \num{-7.2799134134} & & \multicolumn{1}{c}{--} &  & \multicolumn{1}{c}{--} & & \multicolumn{1}{c}{--} & & \multicolumn{1}{c}{--} \\
         \hline \hline
    \end{tabular}
\end{table}

\sisetup{round-mode=places,round-precision=9}

\begin{table}[ht]
        \caption{The convergence of the energy for the 1 $^1S^\texte$ (ground) state of the Be$^{2+}$ ion considering only singlet basis states in the basis set expansion.
    \label{tab:Be2psinglet}}
    \begin{tabular}{@{}l r c r c r c r c r@{}}
         \hline \hline \\[-0.35cm]
         $N$ & \multicolumn{1}{c}{$E_\mathrm{nonrel}$} & & \multicolumn{1}{c}{$E_{\DC}^\pp$}  &&  \multicolumn{1}{c}{$E_{\DC \langle \text{B}\rangle}^\pp$} & &  \multicolumn{1}{c}{$E_{\DC \mathcal{B}_2}^\pp $} &&  \multicolumn{1}{c}{$E_{\DCB}^\pp$} \\ \hline
         100 & \num{-13.65556590956662} & & \num{-13.658257247997} & &\num{-13.657788246} & & \num{-13.6577896010}   & &  \num{-13.6577895946339} \\
         200 & \num{-13.65556622851636} & & \num{-13.6582575962561} & & \num{-13.657788720}  & & \num{-13.657790097} & & \num{-13.6577900901323} \\
         300 & \num{-13.65556623404353} & & \num{-13.6582576022476} & & \num{-13.657788729}  & & \num{-13.6577901066} & &  \num{-13.6577900999711} \\ \hline
         Ref. \cite{drakeHighPrecisionCalculations2006} & \num{-13.655566238} & & \multicolumn{1}{c}{--} &  & \multicolumn{1}{c}{--} & & \multicolumn{1}{c}{--} & & \multicolumn{1}{c}{--} \\
         \hline \hline
    \end{tabular}
\end{table}

\sisetup{round-mode=places,round-precision=10}

\begin{table}[ht]
        \caption{The convergence of the energy for the ground state of the H$_2$ molecule considering only singlet basis states in the basis set expansion.
    \label{tab:H2singlet}}
    \begin{tabular}{@{}l r c r c r c r c r@{}}
         \hline \hline \\[-0.35cm]
         $N$ & \multicolumn{1}{c}{$E_\mathrm{nonrel}$} & & \multicolumn{1}{c}{$E_{\DC}^\pp$}  &&  \multicolumn{1}{c}{$E_{\DC \langle \text{B}\rangle}^\pp$} & &  \multicolumn{1}{c}{$E_{\DC \mathcal{B}_2}^\pp $} &&  \multicolumn{1}{c}{$E_{\DCB}^\pp$} \\ \hline
         \hspace{0.15cm}800 & \num{-1.1744757136} & & \num{-1.1744897537}  &	&	\num{-1.1744866211}  &	&	\num{-1.1744866346}  	& &	\num{-1.17448663459966}  \\
         1000 & \num{-1.1744757136} & & \num{-1.1744897539117} & & \num{-1.1744866214} & & \num{-1.1744866350}  & & \num{-1.17448663493133} \\
         1200 & \num{-1.1744757139} & & \num{-1.17448975396783} & & \num{-1.1744866215} & & \num{-1.1744866351}  & &  \num{-1.17448663505953} \\ \hline
         Ref. \cite{puchalskiRelativisticCorrectionsGround2017a}  & \num{-1.1744757142} & & \multicolumn{1}{c}{--} &  & \multicolumn{1}{c}{--} & & \multicolumn{1}{c}{--} & & \multicolumn{1}{c}{--} \\
         \hline \hline
    \end{tabular}
\end{table}

\clearpage
%

\end{document}